\documentclass[%
reprint,
amsmath,amssymb,
aps,pre,
showkeys]{revtex4-2}
\usepackage{graphicx} 
\usepackage{dcolumn} 
\usepackage{bm} 
\usepackage{float}

\usepackage{amsthm}
\usepackage{xcolor}
\theoremstyle{remark}
\newtheorem{theorem}{Theorem}
\newtheorem{remark}{Remark}
\newtheorem{definition}{Definition}
\newtheorem{lemma}{Lemma}

\newcommand{\abxi}{\boldsymbol{\theta}}

\newcommand{\abxiz}{\boldsymbol{\theta}, {\zeta}}

\newcommand{\F}{F(\omega;\abxiz)}

\newcommand{\Fsq}{F^2(\omega;\abxiz)}

\newcommand{\Tr}{\mathrm{Tr}}
\newcommand{\iu}{\mathrm{i}}
\newcommand{\dd}{\mathrm{d}}

\newcommand{\real}{\mathrm{Re}}

\newcommand{\s}{\sigma}
\newcommand{\sdom}{\{-1,1\}}
\newcommand{\rr}{\mathbb{R}}
\newcommand{\ee}{\mathbb{E}}
\newcommand{\C}{C_{\proximity,\freq}}
\newcommand{\sgn}{\mathrm{sgn}}
\newcommand{\freq}{\kappa}
\newcommand{\scale}{q}
\newcommand{\order}[1]{{O}\left(#1\right)}
\newcommand{\proximity}{\eta}

\newcommand{\Jmatrix}{\boldsymbol{J}}
\newcommand{\leading}{\mathbb{E}[\lambda(\omega)]\bigg|_{\text{lead}}}

\begin{document}
	
	\preprint{APS/123-QED}
	
	\title{Exact power spectra and effective dimensionality in critical block-GOE networks}

	\author{Konstantinos Grammenos}
	\author{Dionissios T. Hristopulos}%

	\affiliation{%
		School of Electrical and Computer Engineering, Technical University of Crete, Chania 73100, Greece
	}

	\date{\today}
	
	\begin{abstract}
		Criticality in neural networks is crucial for optimal information processing. Recent studies have shifted focus from static criticality to frequency-dependent behavior, particularly in oscillatory systems modeled by block-\emph{Gaussian orthogonal ensemble} (block-GOE) matrices. However, analytical tools to predict spectral properties such as power spectrum and effective dimensionality remain scarce. In this work, we derive exact solutions for the network-averaged power spectrum \( \Tr_{N}[\mathbf{S}(\omega)] \), the network autocovariance \( \Tr_{N}[\mathbf{C}(\tau)]\), and the normalized effective dimensionality \( D_{N}(\omega) \) in block-GOE ensembles. These results provide a comprehensive framework for understanding how oscillatory networks operate near criticality, offering insights into the dynamic balance between stability and flexibility in large-scale neural systems.
	\end{abstract}

	\maketitle

	\section{Introduction.} The \emph{critical brain hypothesis} suggests that neural networks operate near a (second-order) \emph{phase transition} to optimize information processing, dynamic range, and adaptability~\cite{PhysRevLett.133.208401, Bak1997-tn, How-critical-is-brainOByrne2022-zl}. At this critical point, the system balances stability and sensitivity, allowing long-range correlations in space and time. While earlier work focused on static, time-averaged properties, recent studies emphasize frequency-dependent neural dynamics~\cite{PhysRevLett.133.208401}, motivated by the role of neural oscillations in coordinating communication across brain regions.
	
	Calvo~\emph{et al.}~\cite{PhysRevLett.133.208401} showed that criticality in the brain is localized within specific frequency bands. Using frequency-dependent \emph{principal component analysis} (PCA), they identified a power-law signature in the power spectrum and a collapse of normalized effective dimensionality at characteristic frequencies, indicating a highly coordinated, traveling-wave state. However, the underlying mathematical mechanism remains descriptive. While the block-\emph{Gaussian orthogonal ensemble} (block-GOE) has been proposed as a model for these oscillatory networks, the analytical relationship between structural connectivity and frequency-dependent spectral properties lacks closed-form expressions.
	
	This work provides the analytical foundation to address this gap. We derive exact closed-form expressions for the first two statistical moments of the \emph{spectral density matrix} (SDM) eigenvalues in the large-$N$ limit for a linear dynamical system with block-GOE connectivity and additive white Gaussian noise. These derivations transform numerical observations of frequency-dependent criticality into a predictive analytical theory.
	
	Our contributions include three key results. First, we derive the \emph{network-averaged power spectrum} (NAPS), $\Tr_{N}[\mathbf{S}(\cdot)]:\mathbb{R}\rightarrow\mathbb{R}_{0,+}$, as a real-valued function of frequency, where $\rr$ denotes the set of real numbers, and $\rr_{0,+}$ denotes the set of nonnegative real numbers. Near the edge of instability, the spectrum develops a singularity (a pole) at a characteristic frequency, following a power-law scaling. Second, using the second statistical moment, we derive the frequency-dependent \emph{normalized effective dimensionality} $D_N(\cdot):\mathbb{R}\rightarrow\mathbb{R}_{0,+}$. We show that as the system approaches the edge of instability, the normalized effective dimensionality collapses, $\lim_{\omega \to \omega_C} \lim_{N\to\infty} D_N(\omega)= 0$, establishing the coordination-criticality link. Finally, we extend our analysis to the time domain by deriving the \emph{network-averaged autocovariance function}, $\Tr_{N}[\mathbf{C}(\cdot)]:\mathbb{R}\rightarrow\mathbb{R}$. We show that the transition from the \emph{subcritical} to the \emph{critical} regime involves a shift from exponential decay to a power-law decay with $\tau^{-1/2}$ dependence, demonstrating \emph{critical slowing down} within the block-GOE framework.
	
	These results provide an analytical framework that advances  the study of the critical brain, offering a tool for characterizing the stability and dynamical range of complex, multi-scale oscillatory networks.

	\section{Theoretical framework.} 
	Neural activity is modeled as a network of $N$ interacting nodes, each with real-valued activity evolving over time~\footnote{In statistical physics, equations of motion that are driven by deterministic drift and stochastic terms are known as Langevin equations.}. The dynamics for all $t\in\mathbb{R}$ are described by a linear system with additive noise~\cite{PhysRevLett.133.208401}:
	\begin{equation}
		\dot{\textbf{x}}(t) = -(\mathbb{I} - g\boldsymbol{W}) \, \mathbf{x}(t) + \boldsymbol{\xi}(t) \,, \label{eq:lin-dyn-sys}
	\end{equation}
	under an initial condition \( \mathbf{x}(0)\), where $\mathbf{x}(t)$ is an $N$-component state vector function, \(\boldsymbol{W}\) is an $N\times N$ \emph{interaction matrix}, $g$ is a nonnegative scalar that determines the strength of the interactions, and \( \mathbb{I} \) is the  $N\times N$  identity matrix. Here, the random vector function $\{\boldsymbol{\xi}(t)\}_{t\in\mathbb{R}}$ models the intrinsic noise, which is assumed to be zero-mean, unit variance, Gaussian, spatially and temporally uncorrelated. Formally, $\left\langle\boldsymbol{\xi}(t)\boldsymbol{\xi}(t-\tau)\right\rangle=\mathbb{I} \, \delta(\tau)$, where $\delta$ is the Dirac delta function and $\left\langle\cdot\right\rangle$ denotes the ensemble average over the noise distribution. Thus, Eq.~\eqref{eq:lin-dyn-sys} represents a \emph{multivariate Ornstein--Uhlenbeck process}~\cite{uhlenbeck_theory_1930, gardiner1985hsm, calvo_robust_2026}. The state vector $\{\mathbf{x}(t)\}_{t\in\mathbb{R}}$ describes neural activity, with $x_i(t)$ corresponding to neuron $i$ for all $i=1,\dots,N$. The matrix $\boldsymbol{W}\in\mathbb{R}^{N\times N}$ encodes the contribution of each neuron to others. The \emph{coupling strength} $g$ determines stability: the system is stable for $g\in (0, g_{C})$, where $g_{C}=1/\max \real (\lambda_{\boldsymbol{W}})$, and unstable otherwise. $\max \real (\lambda_{\boldsymbol{W}}) $ denotes the maximum real part of all eigenvalues of $\boldsymbol{W}$.

	This study focuses on the linear system described by the O-U dynamics, that is, Eq.~\eqref{eq:lin-dyn-sys}. Importantly, the full non-linear version of neural dynamics has a real-valued \emph{activation function} in each neuron, modeled by the vector function $\boldsymbol{F}(\mathbf{x}(t))=[F(x_1(t)), \dots, F(x_N(t))]^{\top}$. The nonlinear system  is given by 
	\begin{equation}
		\frac{\dd }{\dd t}x_i(t) = -x_i(t) + g\sum_{j=1}^{N} W_{ij} F(x_j(t)) + \xi_i(t), \label{eq:full-nonlin}
	\end{equation}
	for all $t\in\mathbb{R}$, where $W_{ij}=[\boldsymbol{W}]_{ij}$ is the $i$-th row and $j$-th column of matrix $\boldsymbol{W}$, $i=1, \dots, N$~\cite{Hu2022-kf, PhysRevLett.133.208401, calvo_robust_2026}. Therefore, the process described in Eq.~\eqref{eq:lin-dyn-sys} is a linearization of Eq.~\eqref{eq:full-nonlin}, near the equilibrium $F(x)\approx x$.

	To analyze dynamic stability, we examine the spectral properties of the connectivity matrix $\boldsymbol{W}$ rather than performing a direct stability analysis of the full nonlinear system in Eq.~\eqref{eq:full-nonlin}~\cite{lyapunov1992general}. In the presence of a bounded activation function such as $F(x) = \tanh(x)$~\cite{Hu2022-kf}, the linear stability regime remains bounded by $g < g_C$. In the unstable regime ($g \ge g_C$), the saturating nonlinearity prevents the unit activations $x_i(t)$ ($i=1,\dots, N$) from diverging to infinity.
	
	For the linearized dynamics studied throughout this work, that is, Eq.~\eqref{eq:lin-dyn-sys}, we restrict the coupling parameter to $g \in [0, g_C)$ and pay particular attention to the limit $g \to g_C^{-}$. This condition guarantees the stationarity of the stochastic process $\{\mathbf{x}(t)\}_{t \in \mathbb{R}}$, ensuring that all statistical observables derived from the spectral density matrix are well-defined. While stationary linear response theory requires $g < g_C$, we note that well-behaved statistical quantities can still be formulated in the nonlinear system above $g_C$. Following standard convention, we refer to the boundary $g \to g_C^{-}$ as the \emph{edge of instability}~\cite{PhysRevLett.133.208401}.
	
	The edge of instability in Eq.~\eqref{eq:lin-dyn-sys} acts as a dynamical \emph{phase transition}, marking a qualitative shift in the global macroscopic state from stable, noise-driven fluctuations to self-sustained chaotic or oscillatory dynamics. Accordingly, we adopt standard nomenclature throughout this work. Specifically, the stable regime $g < g_C$ is termed \emph{subcritical}, the unstable regime $g > g_C$ is \emph{supercritical}, and the boundary $g \approx g_C$ is \emph{critical}.

	Recent work highlights the importance of resolving the frequency dependence, which can help to  reveal emergent properties of the system near an instability such as synchronization~\cite{PhysRevLett.133.208401}.  The classic two-population Wilson--Cowan framework  is a  macroscopic representation of a brain network comprising interacting excitatory ($E$) and inhibitory ($I$) populations that conforms to Eq.~\eqref{eq:lin-dyn-sys}. The coupling of the two populations is described by a $2\times 2$ matrix of the form 
	\begin{equation}
		\label{eq:wc-mat}
		\boldsymbol{W}_{\text{macro}} = \begin{bmatrix}
			w_{EE} & -w_{EI} \\
			w_{IE} & -w_{II}
		\end{bmatrix},
	\end{equation}
	where $w_{EE}, w_{EI}, w_{II}, w_{IE}  > 0$.  The elements of the first column of matrix~\eqref{eq:wc-mat} represent self-excitation and cross-population excitation, while the elements of the second column represent feedback inhibition and self-inhibition terms, respectively~\cite{wilson_mathematical_1973}. Depending on the parametrization, the Wilson-Cowan model can exhibit different types of behavior including transitions between quiescent and oscillatory regimes.
	
	Inspired by the Wilson-Cowan model~\cite{wilson_mathematical_1973}, Calvo \emph{et al.} introduced a connectivity structure with two populations of neurons, excitatory and inhibitory~\footnote{The connectivity type we cite in Eq.~\eqref{eq:w} is not strictly excitatory-inhibitory due to the sign of $\Jmatrix$ elements.}. 
	Their model uses a high-dimensional setting where each subpopulation contains $N/2$ neurons by replacing the scalar weights in the Wilson-Cowan formulation with a shared random matrix $\Jmatrix$, which represents the disordered, isotropic background synaptic connections of the neural medium.
	The interaction matrix $\boldsymbol{W}$ is then defined as follows
	\begin{equation}
		\boldsymbol{W}\triangleq \begin{bmatrix}
			\alpha \Jmatrix & -\beta \Jmatrix \\
			\beta \Jmatrix & \alpha \Jmatrix
		\end{bmatrix}, \label{eq:w}
	\end{equation}
	where $\Jmatrix$ is a Gaussian orthogonal ensemble (GOE) matrix with diagonal elements,  $J_{ii}\sim \mathcal{N}(0,1/N)$, and off-diagonal elements, $J_{ij}\sim\mathcal{N}(0,2/N)$, for $i,j=1,\dots,N/2$, $i\neq j$. The parameters $\alpha, \beta \in\mathbb{R}$ are assumed nonzero unless otherwise stated. In addition, the threshold to instability for this connectivity structure is set by $g_C=1/(2\alpha)$~\cite{PhysRevLett.133.208401}.
	
	The interaction parameter $\alpha > 0$ governs the self-coupling blocks within each subpopulation ($E\to E$ and $I\to I$), while $\beta > 0$ scales the cross-population interactions. Crucially, the antisymmetric block arrangement ($\pm \beta \Jmatrix$) preserves the essential E-I negative feedback loop: the block $+\beta \Jmatrix$ represents the feedforward excitatory drive from $E$ to $I$, whereas the block $-\beta \Jmatrix$ provides the corresponding inhibitory feedback. Under our dimensionless parameterization, $\eta = g/g_C \in [0, 1)$ quantifies the global proximity to the Hopf instability threshold, while $\kappa = \beta/\alpha \ge 0$ isolates the relative strength of cross-population E-I feedback.

	Under the condition of stability, the vector process $\{\mathbf{x}(t)\}_{t\in\mathbb{R}}$ that satisfies Eq.~\eqref{eq:lin-dyn-sys} is a zero-mean, stationary Gaussian process in the steady state (i.e., after the impact of initial conditions has been erased from the system). The respective \emph{covariance function matrix} (CFM) for a specific \(\boldsymbol{W}\) configuration is defined as follows for all $t\in\mathbb{R}$
	\begin{equation}
		\label{eq:cfm}
		\mathbf{C}(\tau)\triangleq \left\langle\mathbf{x}(t) \,\mathbf{x}(t+\tau)^\top\right\rangle \,,
	\end{equation}
	where \(\top\) denotes the transpose and \( \left\langle \cdot \right\rangle\) is the stochastic average over the noise in the steady  state of the system where \( \left\langle \mathbf{x}(t)\right\rangle = \mathbf{0}\)~\footnote{For brevity we use the same symbol for the average over the steady state. The steady-state average differs from the ensemble noise average used for  \(\boldsymbol{\xi}(t)\)}. Given the time invariance of the system~\eqref{eq:lin-dyn-sys}  
	the CFM depends only on the time lag $\tau\in\mathbb{R}$. To analyze neural activity across frequencies, dynamics are studied in the frequency domain~\cite{PhysRevLett.133.208401, Hu2022-kf}. 
	The \emph{spectral density matrix} (SDM) is defined as the power spectrum of the process, expressed via the Wiener-Khinchin theorem as follows
	\begin{equation}
		\mathbf{S}(\omega) = (\boldsymbol{A}+\iu\omega \mathbb{I})^{-1}(\boldsymbol{A}^\top-\iu\omega \mathbb{I})^{-1}, \label{eq:spectral-density}
	\end{equation}
	for all $\omega \in \mathbb{R}$, where $\boldsymbol{A}\triangleq -(\mathbb{I} - g\boldsymbol{W})$, and $\iu$ is the imaginary unit, i.e., $\iu^2=-1$. All frequencies $\omega$ are considered to be non-dimensional, normalized by the neuron response time $\tau_C^{-1}$ (herein we assume $\tau_C=1$). Physical frequencies (in rads/sec) are recovered as $\omega_{\text{phys}}=\omega/\tau_C$.

	From Eq.~\eqref{eq:spectral-density}, $\mathbf{S}(\cdot)$ is a function of $\omega\in\mathbb{R}$, which deterministically depends on $g\in[0,g_C)$, and the sample of the matrix $\boldsymbol{W}$. Therefore, given the random nature of $\boldsymbol{W}$, $\mathbf{S}(\omega)$ is implicitly a random matrix; this implies that the eigenvalues of $\mathbf{S}(\omega)$ are random variables. Importantly, in the large-$N$ limit, the \emph{limiting spectral distribution} (LSD) of SDM eigenvalues follows a probability law determined by the probability density function $\rho(\lambda;\omega)$. In this study, we use the operator $\mathbb{E}[\cdot]$ to denote the \emph{expectation over the limiting spectral distribution of the SDM eigenvalues}, for example, 
	\begin{equation}
		\label{eq:sdm-average}
		\mathbb{E}[\lambda(\omega)]= \int_{D_{\rho,\omega}} \rho(\lambda;\omega) \, \lambda \, \dd \lambda,
	\end{equation}
	\noindent where $D_{\rho,\omega}$ is the support of the eigenvalues at frequency $\omega\in\mathbb{R}$. 
	
	The LSD is analytically available in literature for the oscillatory block-GOE, GOE and  i.i.d. Gaussian connectivity cases~\cite{PhysRevLett.133.208401, Hu2022-kf}. In all cases, while the support of $\rho$ is compact in the subcritical regime, at the edge of instability the right boundary of the support tends to infinity. In addition, the LSD displays power-law dependence of the form $\rho(\lambda; \omega)\sim \lambda^{-5/3}$ in the i.i.d. Gaussian case and $\rho(\lambda; \omega)\sim\lambda^{-7/4}$ in the GOE and block-GOE cases~\cite{PhysRevLett.133.208401,Hu2022-kf}. Instead of the full SDM density, low-order moments such as the ones defined below are often used. 
	
	\medskip 
	\begin{definition}[NAPS]
		The network-averaged power spectrum $\Tr_{N}[\mathbf{S}(\omega)]$ is defined as:
		\begin{equation}
			\Tr_{N}[\mathbf{S}(\omega)]\triangleq \frac{\Tr[\mathbf{S}(\omega)]}{N}, \label{eq:naps-def}
		\end{equation}
		for all $\omega\in\mathbb{R}$. It represents the average trace of $\mathbf{S}(\omega)$, which is equivalent to the average of its eigenvalues. 
		
	\end{definition}
	
	\begin{remark}
		In the large-$N$ limit, NAPS becomes equal to the first statistical moment of $\lambda(\omega)$ over the limiting spectral distribution of $\mathbf{S}(\omega)$, that is, 
		\begin{equation}
			\lim\limits_{N\rightarrow\infty} \Tr_{N}[\mathbf{S}(\omega)]=\mathbb{E}[\lambda(\omega)], \label{eq:naps-ee}
		\end{equation}
		\noindent for all $\omega\in\mathbb{R}$, where the expectation is defined in Eq.~\eqref{eq:sdm-average}.   
		
		The limit statement holds for both connectivity structures we consider (i.i.d. Gaussian and block GOE) and follows from the \emph{self-averaging property} of the  $\boldsymbol{W}$ matrices~\cite{Meckes2017,Guionnet2000-cb}.
		
	\end{remark}

	\medskip 
	
	\begin{definition}[ED]
		The effective dimensionality $D(\omega)$ is defined as:
		\begin{equation}
			D(\omega) = \frac{\left(\sum_{j=1}^{N}\lambda_j(\omega)\right)^2}{\sum_{j=1}^{N}\lambda_j^2(\omega)}, \label{eq:effective-dim-def}
		\end{equation}
		for all $\omega\in\mathbb{R}$, where $\lambda_{1}(\omega), \dots, \lambda_{N}(\omega)$ are the SDM eigenvalues~\cite{PhysRevLett.133.208401, clark_dimension_2023}. This fraction is also known as the participation ratio of the SDM eigenvalues. 
	\end{definition}
	
	Since $D(\omega)$ takes values between $1$ and $N$, a direct consequence of the Cauchy-Schwartz inequality, we focus on the \emph{normalized effective dimensionality} $D_{N}(\omega)\triangleq D(\omega)/N \in [0,1]$.
	
	\begin{remark}
		In the large-$N$ limit the normalized effective dimensionality is given by means of the expectation ratio
		\begin{equation}
			\lim\limits_{N\rightarrow\infty}D_{N}(\omega) = \frac{ \mathbb{E}[\lambda(\omega)]^2}{\mathbb{E}[\lambda^{2}(\omega)]}\,,
			\label{eq:effective-dim-large-n-limit}
		\end{equation}
		for all $\omega \in \mathbb{R}$.
	\end{remark}
	
	\medskip
	
	\begin{definition}[NAAC]
		The network-averaged autocovariance is defined as:
		\begin{equation}
			\Tr_N[\mathbf{C}(\tau)]\triangleq \frac{\Tr[\mathbf{C}(\tau)]}{N},
		\end{equation}
		for all $\tau\in\mathbb{R}$. In the large-$N$ limit, this equals the Fourier transform of the first statistical moment of the random eigenvalue $\lambda(\cdot)$ over the limiting spectral distribution of $\mathbf{S}(\omega)$, as a function of the cyclical frequency $\omega\in\rr$. That is, in the large-$N$ limit, NAAC is equal to the Fourier transform of Eq.~\eqref{eq:naps-ee}.
	\end{definition}
	
	\medskip
	
	In linear i.i.d. Gaussian networks, the NAPS, ED, and NAAC are studied in~\cite{Hu2022-kf}. Specifically, it is found that
	\begin{equation}
		\lim_{N \to \infty} \Tr_{N}[\mathbf{S}(\omega)]=\frac{1}{1-g^2+\omega^2\tau_C^2}, \label{eq:net-averaged-iid}
	\end{equation}
	for all $\omega\in\mathbb{R}$~\cite{Hu2022-kf}. Respectively, in the time domain it holds
	\begin{equation}
		\lim_{N \to \infty} \Tr_{N}[\mathbf{C}(\tau)]=\frac{1}{2\sqrt{1-g^2}\tau_C}e^{-{\sqrt{1-g^2}|\tau|}/{\tau_C}}, \label{eq:corr-net-averaged-iid}
	\end{equation}
	for all $\tau\in\mathbb{R}$. In Eqs.~\eqref{eq:net-averaged-iid}-\eqref{eq:corr-net-averaged-iid}, $\tau_C$ is a positive constant that characterizes the neurons' response time to signals. In this work, we always work for the case $\tau_C=1$. In addition, $g$ represents the coupling strength: $g\in[0,1)$. Notably, Eqs.~\eqref{eq:net-averaged-iid}-\eqref{eq:corr-net-averaged-iid} imply that near the edge of instability, the Fourier duality of these quantities breaks. Specifically, the fact that $\lim_{N \to \infty} \Tr_{N}[\mathbf{S}(\omega)]\propto\omega^{-2}$ implies that $\Tr_{N}[\mathbf{C}(\tau)]\propto |\tau|$. This does not match the behavior of Eq.~\eqref{eq:corr-net-averaged-iid}, since at the edge of instability the right-hand side diverges to infinity. As we subsequently demonstrate, such a breakdown does not occur in the block-GOE network.
	As for normalized ED, in i.i.d. Gaussian networks it is given by
	\begin{equation}
		\lim_{N \to \infty} D_{N}(\omega)=\left(1 - \frac{g^2}{1+\omega^2}\right)^2,
	\end{equation}
	for any $\omega\in\mathbb{R}$~\cite{Hu2022-kf}.

	\vspace{1 cm}

	\section{Theoretical contributions}\label{sec:theoretical_contrib}
	We derive closed-form expressions for $\Tr_{N}[\mathbf{S}(\omega)]$, $\Tr_{N}[\mathbf{C}(\tau)]$, and $D_{N}(\omega)$ for block-GOE networks in the large-$N$ limit. The calculations are based on the first two statistical moments of the SDM eigenvalues over the limiting spectral distribution.
	
	First, note that the parametrization of Eqs.~\eqref{eq:lin-dyn-sys} and~\eqref{eq:w} is redundant, because the network connectivity is controlled by three parameters, \(\alpha, \beta, g\). We rescale the system by defining  the normalized coupling strength $\proximity \triangleq g/g_C=2\alpha g$. By definition, in the stable regime it holds that $\proximity \in [0,1)$. We also define $\freq \triangleq \beta/\alpha$. Then, for all $\alpha\neq0$, the product \(g \boldsymbol{W}\) of the coupling strength and the interaction matrix is expressed as follows in terms of two parameters
	\begin{align}
		g\boldsymbol{W} &= \frac{\proximity}{2}\begin{bmatrix}
			\Jmatrix & - \freq \Jmatrix \\
			\freq \Jmatrix & \Jmatrix
		\end{bmatrix}\,. \label{eq:new-W}
	\end{align}

	Therefore, based on Eq.~\eqref{eq:lin-dyn-sys} and Eq.~\eqref{eq:new-W}, without loss of generality the system can be expressed in terms of $\freq$, which can be limited to $\freq \geq 0$  and $\proximity \in [0,1)$. More information regarding the change of variables described above can be found in Appendix~\ref{app:notation}. 
	
	The parameter $\freq$ controls the interactions between the two populations and determines the oscillatory structure of the network (in dimensionless units), while $\proximity$ measures the \emph{proximity to instability}. The \emph{characteristic frequency} is $\omega_C=\pm \proximity\freq$, which ranges from zero for $\eta=0$ to $\pm\freq$ for $\eta \to 1^{-}$. Hence,  $\freq$ represents the \emph{critical cyclic frequency} of the network. 
	Our results are derived in terms of the parameter vector \( \boldsymbol{\theta} \triangleq \left(\proximity, \ \freq, \ \sigma \right)^\top,\) where  $\sigma\in\{-1,1\}$ is a sign variable that distinguishes between two branches of SDM eigenvalues obtained from the same eigenvalue of \(\boldsymbol{J}\) (see Appendix~\ref{app:notation}). 
	The moments of the eigenvalue distribution are obtained by summation  over both branches. 
	As we discuss below, the first- and second-order moments of the eigenvalue distribution are invariant under the time reversal transformation $\freq \to -\freq$. Hence, in the following we restrict $\freq>0$ without loss of generality and investigate the case $\freq=0$ separately in Appendix~\ref{app:rho-0}. 
	
	\medskip 
	
	\subsection{Network-averaged power spectrum} 
	
	This section focuses on the calculation of the normalized SDM trace, $\Tr_{N}[\mathbf{S}(\omega)]$, in the large-$N$ limit for  ensembles with block-GOE interaction matrices. 
	
	\begin{theorem}[NAPS closed form]
		\label{theorem:NAPS} 
		Let  $\lambda(\omega)$  denote the random  SDM eigenvalues for the spectral matrix~\eqref{eq:spectral-density} with interaction matrix $\boldsymbol{W}$ given by Eq.~\eqref{eq:w}. Then, in the large-$N$ limit, the expectation of $\lambda(\omega)$ over the limiting spectral distribution of $\mathbf{S}(\omega)$ is given by
		\begin{equation}
			\label{eq:naps_closed}
			\mathbb{E}[\lambda(\omega)]=\frac{1}{2}\sum_{\sigma\in\{-1,1\}} \Lambda_1(\omega;\boldsymbol{\theta}),
		\end{equation}
		for all $\omega\in\mathbb{R}$.  For fixed $\boldsymbol{\theta}$,  $\Lambda_1(\cdot\,;\boldsymbol{\theta}):\mathbb{R}\rightarrow\mathbb{R}$ is a real-valued function that represents the SDM eigenvalue expectation per branch. It is given by the following equation
		\begin{widetext}
			\begin{align}
				\Lambda_1(\omega;\boldsymbol{\theta}) = 
				\frac{\sqrt{2}}{\proximity^2(1+\freq^2)A(\omega;\freq,\sigma)}
				\left[ 
				\sqrt{H^2(\omega;\abxi) + 4A^2(\omega;\freq,\sigma) B^2(\omega;\freq,\sigma)} - H(\omega;\abxi)
				\right]^{1/2} - \frac{2}{\proximity^2(1+\freq^2)},
				\label{eq:first-moment}
			\end{align}
		\end{widetext} 
		for all $\omega \in \rr \setminus \{-\freq, \freq \} $, $\proximity\in[0,1)$, and $\s \in \sdom$. The spectral function $H(\omega;\abxi)$ in Eq.~\eqref{eq:first-moment}  is defined as
		\begin{equation}
			H(\omega;\abxi) = B^2(\omega;\freq,\s) - A^2(\omega;\freq,\s) - 4\C^2,
			\label{eq:H}
		\end{equation}
		where $\C\triangleq \proximity(1+\freq^2)/2$. The spectral branch functions $A(\omega;\freq,\s), B(\omega;\freq,\s)$ are  defined as follows:
		\begin{equation}
			A(\omega;\freq,\s) = |\freq+\s \omega|, \ B(\omega;\freq,\s)= \s \freq \omega-1\,.
			\label{eq:paramsABC}
		\end{equation}
	\end{theorem}
	
	The proof of Theorem~\ref{theorem:NAPS} is given in Appendix~\ref{app:naps}.
	
	\medskip

	The function $A(\omega;\freq,\s)$ measures the absolute distance between the signed frequency $\omega$ and the critical frequency,  $\freq$, of the network. Naturally, $A(\omega;\freq,\s)$ is equal to zero if and only if $\omega=-\s\freq$. This creates a NAPS singularity, both in the subcritical regime ($\proximity<1$) and at the edge of instability ($\proximity\to1^-$). However, for $\proximity<1$, the singularity is removable, and the limit of $\Lambda_1(\omega;\abxi)$ as $\omega\to-\s \freq$ is finite.  At the same time, at $\proximity \to 1^-$, NAPS has an algebraic singularity (an infinity).

	\begin{lemma}[NAPS singularity at $\omega_C$]
		\label{lemma:naps_singular}
		Consider the network-averaged power spectrum as determined by Eqs.~\eqref{eq:naps_closed} and~\eqref{eq:first-moment}.  
		At the critical connectivity limit, $\proximity\to1^-$, NAPS develops an algebraic singularity at the characteristic frequency $\omega_{C}=\pm\freq$.  In particular,
		\begin{equation}
			\mathbb{E}[\lambda(\omega)] \sim \scale_{\pm}(\freq) \, |\omega -\omega_{C}|^{\nu}, \label{eq:sing-power-law}
		\end{equation}
		with $\nu=-1/2$, as $\omega \to \omega_{C}$ where $\omega_{C} \in \{ -\freq, \freq\}$. In Eq.~\eqref{eq:sing-power-law}, $\scale_{\pm}(\freq)$ is defined as
		\begin{equation}
			\scale_{\pm}(\freq) = \frac{\left(\sqrt{1+\freq^2} \mp \freq\right)^{1/2}}{\left( 1+\freq^2\right)^{1/2}}, \label{eq:q-scale}
		\end{equation}
		for all $\freq>0$. The function $\scale_{\pm}(\freq)$  involves two branches, one for each side of the transition, where the positive sign corresponds to $|\omega|>\freq$ and the negative to $|\omega|<\freq$.
	\end{lemma}

	\smallskip The proof is given in Appendix~\ref{app:behavior-close-to-char}.  Relevant details can be found in Eq.~\eqref{eq:crit-asymp} and Eq.~\eqref{eq:crit-asymp-scale}.
	
	\smallskip
	
	At the characteristic frequency  $\omega_C \in \{-\proximity\freq, \proximity\freq\}$, NAPS gets infinite only for $\proximity\to 1^{-}$. For $\proximity<1$, NAPS has a smooth peak at $\omega_C$, where $\omega_C \in \{-\proximity\freq, \proximity\freq\}$. 
	In the subcritical regime, $\eta < 1$, NAPS is bounded near $\omega_C$. The law of $|\omega-\omega_C|^{-1/2}$ of NAPS holds only as an intermediate asymptotic, for $1-\eta \ll |\omega-\omega_C|\ll 1$. More details are given in Appendix~\ref{app:behavior-close-to-char}.

	\begin{figure}[tb] 
		\centering
		\includegraphics[width=0.98\linewidth]{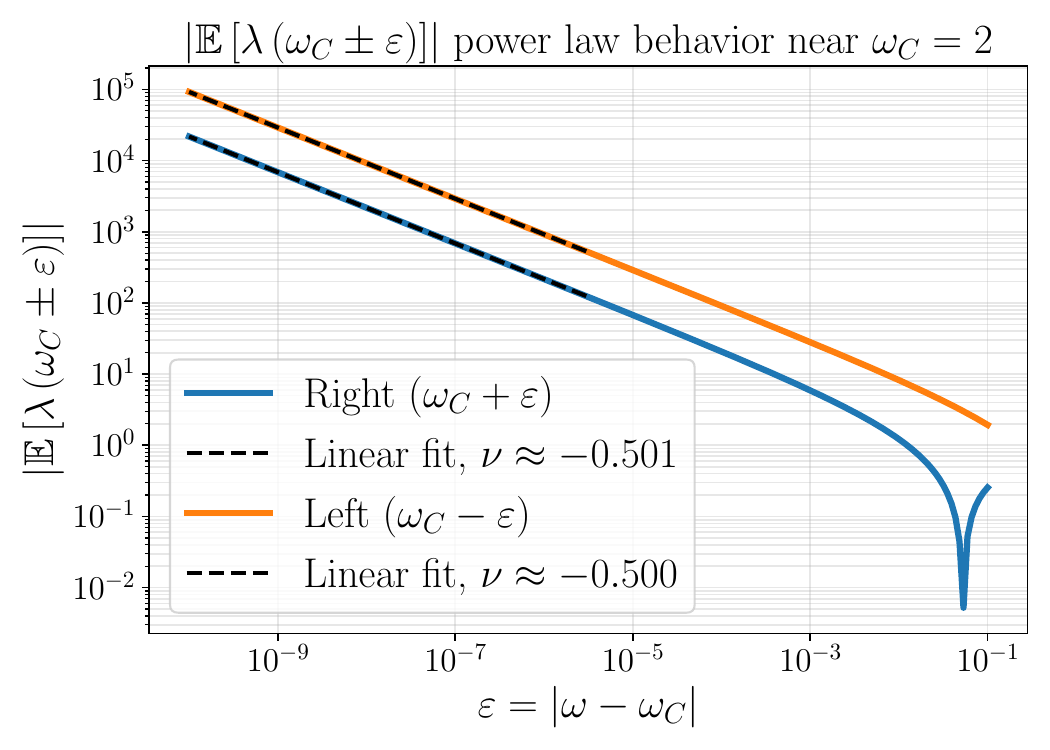}
		\caption{Plot of $\mathbb{E}[\lambda(\omega)]$ at $\proximity\to1^-$ against $|\omega-\omega_C|$, at each side of the singularity $\omega_C=2$.}
		\label{fig:naps-singularity}
	\end{figure}
	
	From this point on, we use the abbreviation LHS to denote the \emph{left hand side} and RHS to denote the \emph{right hand side}. In Fig.~\ref{fig:naps-singularity} we show the dependence of NAPS versus $\varepsilon=|\omega-\omega_C|$ based on Eqs.~\eqref{eq:naps_closed} and~\eqref{eq:lambda1-closed} in the critical regime \(\proximity=1\). The curves  shown in the doubly-logarithmic plots (blue line for \(\omega=\varepsilon + \omega_C\)  and orange line for \(\omega= \omega_C -\varepsilon\)) are obtained for $\freq=2$. The cyclical frequency is sampled at 200 equidistant points in log space over the interval $[\omega_C+10^{-10}, \omega_C +10^{-1}]$ (corresponding to frequencies on the RHS of \(\omega_C\)) and an equal number of points over the interval $[\omega_C-10^{-10}, \omega_C-10^{-1}]$ (corresponding to frequencies on the LHS of \(\omega_C\)). To compare with the expected singular scaling near $\omega_C=2$, Fig.~\ref{fig:naps-singularity} also shows least-squares fits of the NAPS values against the power-law function  $f(\varepsilon) = \scale_{\pm} \, |\varepsilon|^{\nu}$ for the two branches around \(\omega_C\). The linear (due to the double logarithms) fits are restricted to 100 points of each frequency interval that are closer to \(\omega_C\), where the NAPS curves approximate straight lines. In the right branch ($\omega > \omega_C$), the fit produces an exponent $\nu_{+} = -0.5005 \pm 3.39 \times 10^{-5}$ and an amplitude $\scale_+ = 0.2149$, with a coefficient of determination $R^2 = 1.00$. Similarly, on the left branch ($\omega < \omega_C$), we obtain $\nu_{-} = -0.5001 \pm 7.99 \times 10^{-6}$ and $\scale_- = 0.918$ ($R^2 = 1.00$). Both fitted exponents are in excellent agreement with the theoretical prediction $\nu = -1/2$, confirming the expected $\mathbb{E}[\lambda(\omega)] = \order{|\omega - \omega_C|^{-1/2}}$ divergence near the transition~(see Lemma~\ref{lemma:naps_singular}). The same holds for the scaling factors $\scale_{\pm}$, where the right fit $\scale_+ = 0.2149$ (see above) approximates the theoretical value $\scale_+(2)\approx 0.217$, and, similarly, the fitted $\scale_{-} = 0.918$ approximates the theoretical $q_{-}(2)=0.920$.  
	
	\begin{lemma}[NAPS behavior at infinity]
		\label{lemma:naps_infty}
		Consider the network-averaged power spectrum as determined by Eqs.~\eqref{eq:naps_closed} and~\eqref{eq:first-moment}. The asymptotic behavior at $\omega \to \infty$ is given by
		\begin{equation}
			\ee[\lambda(\omega)] = \frac{1}{\omega^2} +  \order{\omega^{-4}},
		\end{equation}
		for all $\proximity\in[0,1)$ and $\freq > 0$.
	\end{lemma}
	
\begin{figure}[tb] 
		\centering
		\includegraphics[width=0.98\linewidth]{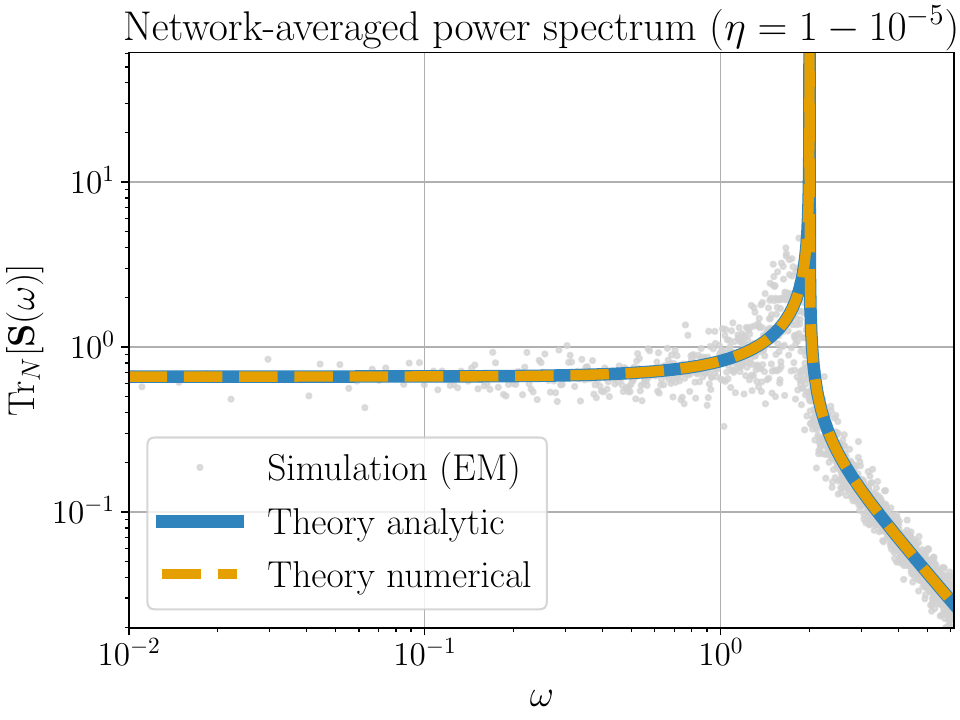}
		\caption{NAPS as a function of $\omega$ for $\freq=2$ and $\proximity=1-10^{-5}$. Blue (solid): Closed-form theoretical solution (see Eq.~\eqref{eq:first-moment}). Orange (dashed): Numerical integration of eigenvalues over the limiting spectral distribution of the SDM. Blue (dots): Euler-Maruyama simulation. Read more information in the paragraph below.}
		\label{fig:naps}
	\end{figure}

	The behavior of NAPS as a function of frequency is illustrated in Figure~\ref{fig:naps}, for $\freq=2$ and $\proximity=1-10^{-5}$. The figure captures three results: (1) the analytical expression of NAPS given in Theorem~\ref{theorem:NAPS} (continuous line, blue online); (2) a numerical NAPS function, derived by integrating Eq.~\eqref{eq:h-integral} using numerical quadrature.  (broken line, orange online); (3) Numerical NAPS estimates evaluated numerically on a logarithmic grid of $n=10 \times 10^{3}$ discrete frequencies, whose consecutive values differ by a fixed multiplicative factor, spanning $[10^{-2}, 1.1\omega_C]$. The estimates have been derived by solving Eq.~\eqref{eq:lin-dyn-sys} using the Euler-Maruyama integration method~\cite{euler_maruyama}. A network with $N=50$ nodes is used and a time step of $\Delta t=0.02$ for a total duration of $T=1.7\times 10^3$ (to achieve the steady state), and a leading burn-in $T=100$. The NAPS is then calculated as follows. After discarding the initial transient, the trajectory of each node $x_i(t)$ is Fourier-transformed in time using the discrete Fourier transform, with the result rescaled by $\Delta t$ to approximate the continuous transform. The squared modulus of this transform, divided by the observation time $T$, gives the power spectrum of each node. These single-node spectra are then averaged over all $N$ nodes, which yields the NAPS~(see Eq.~\eqref{eq:naps-def}). Finally, the result is evaluated on the logarithmic frequency grid described above. All three methods show very good agreement, thus confirming the validity of the analytical expressions.

	\subsection{Normalized effective dimensionality} 
	
	This section focuses on the calculation of the normalized effective dimensionality,  $D_N(\omega)$, in the large-$N$ limit. 
	
	\begin{theorem}[Second moment of eigenvalues distribution]
		\label{theorem:ED}
		Consider the spectral density matrix $\mathbf{S}(\omega)$ defined in Eq.~\eqref{eq:spectral-density}.  Then, in the large-$N$ limit the expectation of the second moment of the SDM random eigenvalues $\lambda(\omega)$ is given by means of the expression
		
		\begin{align}
			&\mathbb{E}[\lambda^2(\omega)]  =
			\sum_{\sigma \in \{-1,1\}}\frac{1}{2 A^2(\omega;\freq,\s)} \notag \\
			&\qquad\times\left[
			\frac{2\pi}{\proximity^2} 
			\left(\sum_{\zeta\in\{-1,1\}} G(\omega;\abxi,\zeta)\right)
			+ \frac{1+\freq^2}{2}\Lambda_1(\omega;\abxi)
			\right], \label{eq:second-moment}
		\end{align}
		for all $\omega\in\mathbb{R} \setminus \{-\freq,\freq\}$, where $A(\omega;\freq,\s)$ is as defined in Eq.~\eqref{eq:paramsABC}. The real-valued spectral function  $G(\omega;\abxi, \zeta)$ is given by  
		\begin{align}
			G(\omega;\abxi,\zeta)&=\frac{\sqrt{(4\pi)^2+\Fsq} - \F}{4\pi \sqrt{(4\pi )^2+\Fsq}}, 
			\intertext{where} 
			\F &= \frac{2\pi}{\C} \left[ A(\omega;\freq,\s) + \iu \, \zeta B(\omega;\freq,\s) \right],
		\end{align}  
		for all $\omega\in\mathbb{R} \setminus \{-\freq,\freq\}$, 
		$\proximity\in[0,1)$, and $\zeta\in\{-1,1\}$, where $\zeta$ is an inner sign variable.    
	\end{theorem}
	
	The proof is given in Appendix~\ref{app:ED}.
	
	\medskip

	Theorems~\ref{theorem:NAPS} and~\ref{theorem:ED} contain all the information needed to express in closed form the large-$N$ limit of the normalized effective dimensionality as defined in Eq.~\eqref{eq:effective-dim-large-n-limit}. 
	\medskip

	\begin{lemma}[NED collapse near $\omega_C$]
		\label{lemma:dim-collapse}
		Consider the normalized effective dimensionality in the large-$N$ limit,
		obtained via Eq.~\eqref{eq:effective-dim-large-n-limit} using Eqs.~\eqref{eq:first-moment} and \eqref{eq:second-moment}. At the edge of instability, $\eta \to 1^-$,
		\begin{equation}
			\begin{split}
				\lim_{N\to\infty} D_N(\omega) &\sim d_{\pm}(\kappa)\,|\omega - \omega_C|^{3/2},\\
				d_{\pm}(\kappa) &= \frac{4\left(\sqrt{1+\kappa^2} \pm \kappa\right)^{1/2}}{1+\kappa^2},
			\end{split}
			\label{eq:NED_collapse}
		\end{equation}
		as $\omega \to \omega_C$, where $\omega_C \in \{-\kappa, \kappa\}$. The upper sign
		corresponds to $|\omega| > |\omega_C|$ and the lower sign to
		$|\omega| < |\omega_C|$. Consequently,
		\begin{equation}
			\lim_{\omega \to \omega_C}\, \lim_{\eta \to 1^-}\, \lim_{N\to\infty} D_N(\omega) = 0.
			\label{eq:NED_zero}
		\end{equation}
		For fixed $\eta < 1$, $\lim_{N\to\infty} D_N(\omega)$ is positive near
		$\pm\eta\kappa$, and its local minimum there is of order $(1-\eta)^{3/2}$.
	\end{lemma}
	
	The proof is given in Appendix~\ref{app:C3}.

	\begin{figure}[!htb]
		\centering
		\includegraphics[width=0.98\linewidth]{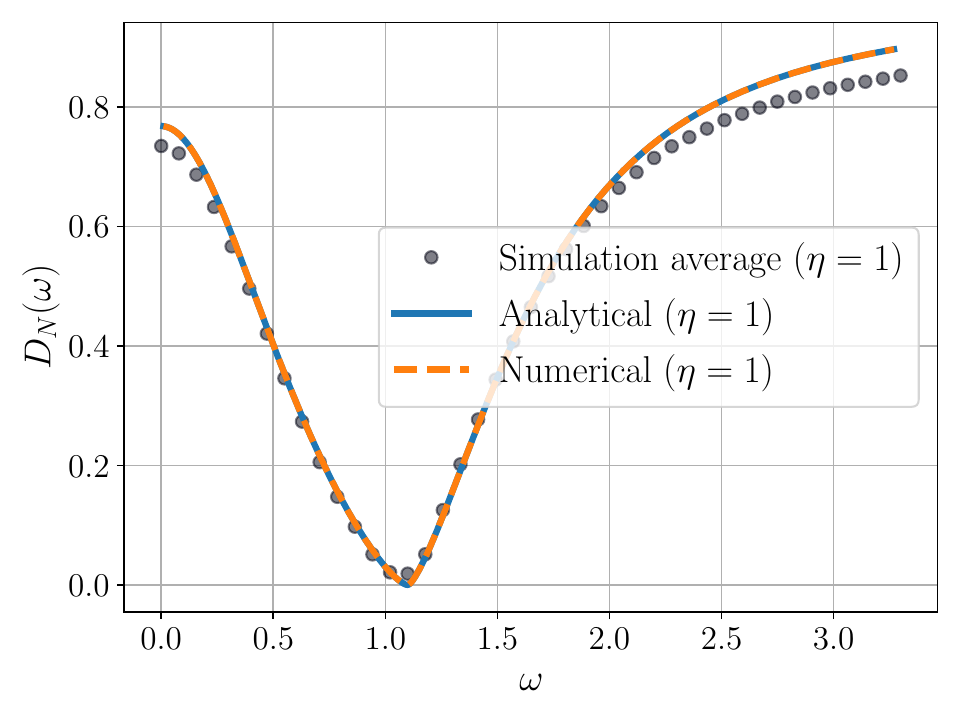}
		\caption{Normalized effective dimensionality for $\freq=1.1$ at the edge of instability, i.e., $\proximity=1$. Blue (solid): Analytical solution at the  infinite network limit using Eqs.~\eqref{eq:first-moment}, \eqref{eq:second-moment}. Purple (dashed): Numerical estimation based on numerical integration of the eigenvalues (first and second statistical moment) over the limiting spectral distribution of the SDM. Blue markers (dots): Simulation using the Euler-Maruyama method. }
		\label{fig:ED}
	\end{figure}
	
	Figure~\ref{fig:ED} shows the normalized ED for $\proximity=1$ and $\freq=1.1$ as a function of frequency. The blue line is the analytical expression at the infinite network limit, using Eq.~\eqref{eq:effective-dim-large-n-limit} (see also Eq.~\eqref{eq:first-moment} and Eq.~\eqref{eq:second-moment}). The dashed orange line is the second moment of the random eigenvalue integrated over the large-$N$ limit distribution, cf. Eqs.~\eqref{eq:n-mom-a-b-c} and~\eqref{eq:integ-secon-mom}. Lastly, the dark dots represent estimates of the normalized ED for $M=8$ realizations of the dynamics presented in Eq.~\eqref{eq:lin-dyn-sys}. We computed the dynamical evolution of the system using the Euler--Maruyama scheme~\cite{euler_maruyama} with a time step $\Delta t = 0.01$. To ensure that the process reaches its stationary state, we discard the first $4\times10^{3}$ time steps of each realization as burn-in and then record $48 \times10^{3}$ units of time ($4.8\times 10^{6}$ points). We estimate the spectral densities with Welch's method~\cite{welch_use_1967}, dividing each realization into $K = 600$ non-overlapping segments of $8\times10^{3}$ points, tapering each segment with a Hann window, and averaging the resulting periodograms. 
	
	\begin{figure}[tb] 
		\centering
		\includegraphics[width=0.98\linewidth]{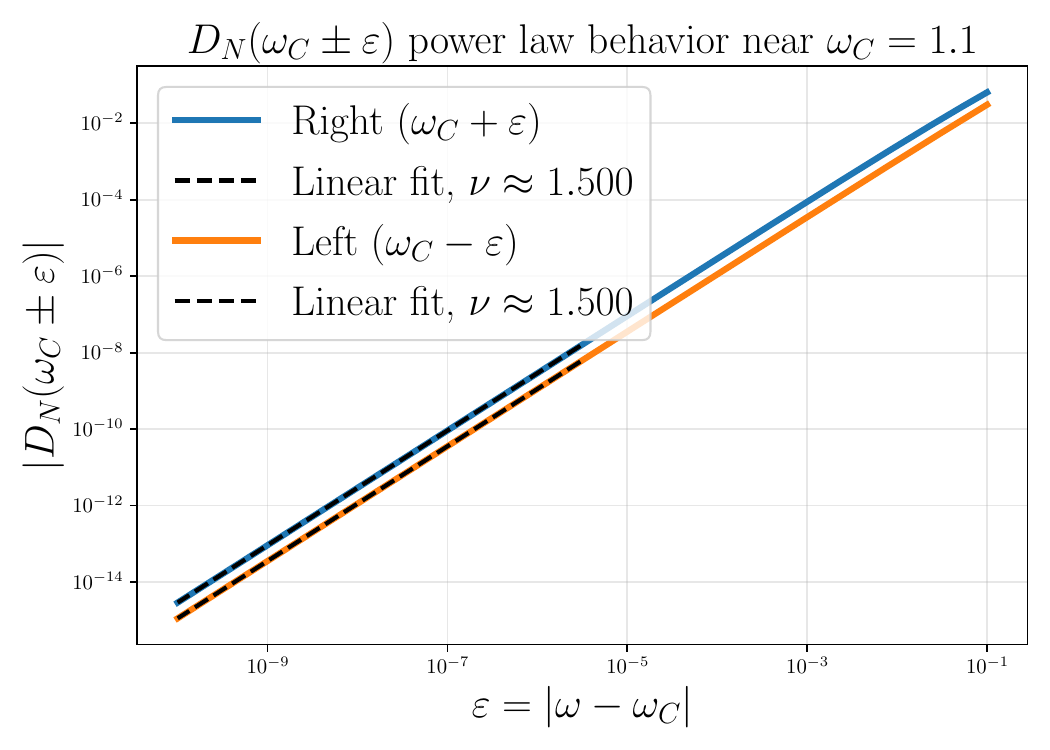}
		\caption{Plot of $D_N(\omega)$ at $\eta\to1^{-}$ against $\pm(\omega-\omega_C)$ at each side of the algebraic zero point $\omega_C.$}
		\label{fig:ED-order}
	\end{figure}
	
	Figure~\ref{fig:ED-order} is a visual validation of Lemma~\ref{lemma:dim-collapse}. To validate the scaling behavior near the algebraic zero point $\omega_C=\freq$, we performed a least-squares fit to the power-law form $f(\omega) = d_{\pm} |\omega - \omega_C|^\nu$ separately on both sides of the transition. The same discretization of the frequency space is used as in Fig.~\ref{fig:naps-singularity}.  On the RHS branch ($\omega > \omega_C$), the fit yields an exponent $\nu_+ = 1.4998 \pm 9.70 \times 10^{-6}$ and amplitude $d_+ = 2.902$, with a coefficient of determination $R^2 = 1.00$. On the LHS branch ($\omega < \omega_{C}$), we obtain $\nu_- = 1.4999 \pm 3.75 \times 10^{-6}$ and $d_- = 1.124$ ($R^2 = 1.00$). Both exponent values are in agreement with the theoretical prediction $\nu = 3/2$ of Lemma~\ref{lemma:dim-collapse}. Additionally, substituting $\freq=1.1$ into the theoretical prediction of $d_\pm(\freq)$ from Lemma~\ref{lemma:dim-collapse}, gives $d_+(1.1)=2.911$ and $d_-(1.1)=1.125$. Therefore, the fitted values are in very good agreement with the theory.

	\medskip
	
	\subsection{Network-averaged autocovariance function} 
	
	This section focuses on the calculation of the normalized trace of the autocovariance function matrix, $\Tr_{N}[\mathbf{C}(\tau)]$, as a function of time delay $\tau \in \rr$. We derive an analytical expression which is valid in the infinite network limit.
	Unlike the frequency-domain metrics, $\lim_{N\to\infty}\Tr_{N}[\mathbf{C}(\tau)]$ is not given by an elementary function; it is defined in terms of the \emph{confluent hypergeometric function of the first kind} of two variables, that is, $\Phi_1(a,b,c; x,y)$, evaluated at $(x,y)$ given parameters $(a,b,c)$~\cite{Abramowitz1964}. We simplify this expression for $\proximity\to1^-$. For both regimes, we present low order asymptotic expansions.
	
	\medskip
	
	\paragraph{Subcritical regime ($\proximity<1$):}
	We present and discuss the exact NAAC expression which is valid for $\proximity<1$ in the large-$N$ limit. 
	
	\begin{theorem}[NAAC subcritical regime]\label{thr:nacc-subcrit}
		In the subcritical regime, that is, for $\proximity<1$, the network-averaged autocovariance function is given by
		
		\begin{widetext}
			\begin{align}
				\lim_{N\to \infty}\Tr_{N}[\mathbf{C}(\tau)]  & = \frac{e^{-(1+\proximity)|\tau|}}{4(1+\proximity)}
				\sum_{\sigma\in\{-1,1\}}e^{\iu 2 \sigma \freq \proximity\tau } \Phi_1\left(\frac{3}{2},1,3;\frac{2\proximity}{1+\proximity},2\proximity(|\tau|-i\sigma \freq \tau)\right),\label{eq:closed-ctau}
			\end{align}
		\end{widetext}
		for all $\tau\in\mathbb{R}$.
	\end{theorem}
	
	\medskip
	The proof of the Theorem~\ref{thr:nacc-subcrit} is given in Appendix~\ref{app:naac-subcritical}.
	
	Notably, the trace of the network variances is given by
	\begin{equation}
		\lim_{N\to \infty}\Tr_{N}[\mathbf{C}(0)] = \frac{1-\sqrt{1-\proximity^2}}{\proximity^2}\,, \label{eq:c-0}
	\end{equation}
	for all $\proximity\in(0,1)$~\cite[see][Ch.~2, Sec.~2.1]{magnus_formulas_1966}. As a result, NAAC at zero lag is equal to $1/2$, when $\proximity \to 0^+$, and $1$, when $\proximity \to 1^-$. 
	
	The asymptotic expansion which we present in Eq.~\eqref{eq:subcrit-asymp} of Appendix~\ref{app:first-moment-eigenvalues-time} shows that the leading term of NAAC in the subcritical regime is of order $\order{|\tau|^{-3/2}e^{-(1-\proximity)|\tau|}}$, for sufficiently large $|\tau|$. In addition, the asymptotic expansion shows that oscillations do not survive the tail. The frequency of $2\freq\proximity$ only shapes NAAC at short and intermediate lags.
	
	\medskip

	\paragraph{Critical regime ($\proximity\to1^{-}$):}
	We present an exact NAAC expression which is valid in the critical regime (i.e., for $\proximity\to1^-$)  and the large-$N$ limit.
	
	\begin{theorem}[NAAC critical regime] \label{thr:naac-crit}
		At the edge of instability, that is, for $\proximity\to1^-$, the network-averaged autocovariance function is given by
		\begin{widetext}
			\begin{align}
				\lim_{N\to \infty}&\Tr_{N}[\mathbf{C}(\tau)]\bigg|_{\proximity \to 1^-} =\frac{e^{-|\tau|}}{2} 
				\sum_{\s \in \sdom} \left[I_0\left(|\tau|- \iu \s \freq \tau\right)+I_1\left(|\tau|- \iu \s \freq \tau\right)\right]
				, \label{eq:closed-form-critical-tau-re}
			\end{align}
		\end{widetext}    
		for all $\tau\in\mathbb{R}$, where $I_0(\cdot),$ $I_1(\cdot),$ denote the modified Bessel functions of the first kind of orders 0 and 1, respectively. 
	\end{theorem}

	\smallskip
	The proof is given in Appendix~\ref{app:naac-critical}.
	
	\medskip
	
	Importantly, NAAC displays a power-law decay of $\tau^{-1/2}$. More specifically, at the long-time limit we obtain the NAAC behavior is determined by the following Lemma~\ref{lemma:crit-decay}. 
	
	\begin{lemma}
		\label{lemma:crit-decay}
		Consider the network-averaged autocovariance function of Eq.~\eqref{eq:closed-form-critical-tau-re}. For sufficiently large time lag, $|\tau|\gg1$,
		\begin{align}
			&\lim_{N\to \infty} \Tr_{N}[\mathbf{C}(\tau)]\bigg|_{\proximity \to 1^-} = \sqrt{\frac{2}{\pi |\tau|}}
			\frac{1}{(1+\freq^2)^{1/4}}
			\notag \\
			&\quad \times \cos\left( \freq |\tau|-\frac{1}{2}\arctan \freq \right) + O(|\tau|^{-3/2}).\label{eq:C-critical-decay}
		\end{align}
	\end{lemma}
	
	\smallskip
	The proof is given in Appendix~\ref{app:asympt-crit}.
	
	\smallskip

	This result shows that critical slowing down emerges together with oscillations of cyclical frequency $\freq$ at the edge of instability. Unlike i.i.d. Gaussian networks, block-GOE networks preserve the Fourier duality between NAAC and NAPS at the edge of instability. The algebraic tail of Eq.~\eqref{eq:C-critical-decay} determines the singular part of NAPS. Since the remainder $\order{|\tau|^{-3/2}}$ is absolutely integrable, the Fourier transform of the leading term of Eq.~\eqref{eq:C-critical-decay} yields $\ee[\lambda(\omega)] \sim \scale_{\pm}(\freq)|\omega-\omega_C|^{-1/2}$, as $\omega\to\omega_C$, $\omega_C \in \{-\freq,\freq\}$, where $\scale_\pm(\freq)$ is the same as in Eq.~\eqref{eq:q-scale}, with positive sign corresponding to $|\omega|>|\omega_C|$. The above result reproduces both the exponent $\nu=-1/2$ and the amplitudes on each branch of the singularity, as derived in Lemma~\ref{lemma:naps_singular}. The above result is proven in Appendix~\ref{app:D2c}. The asymmetry originates in the phase lag $-\arctan(\freq)/2$ of the oscillation in Eq.~\eqref{eq:C-critical-decay}. Without it, $\scale_{-}(\freq)=\scale_{+}(\freq)$, for all $\freq > 0$. Instead, Eq.~\eqref{eq:C-critical-decay} gives $q_-(\freq)/q_+(\freq)=\sqrt{1+\freq^2}+\freq$. The phase lag is therefore the time-domain signature of the spectral asymmetry. 
	
	\medskip
	
	\begin{figure}[tb]
		\centering
		\includegraphics[width=0.98\linewidth]{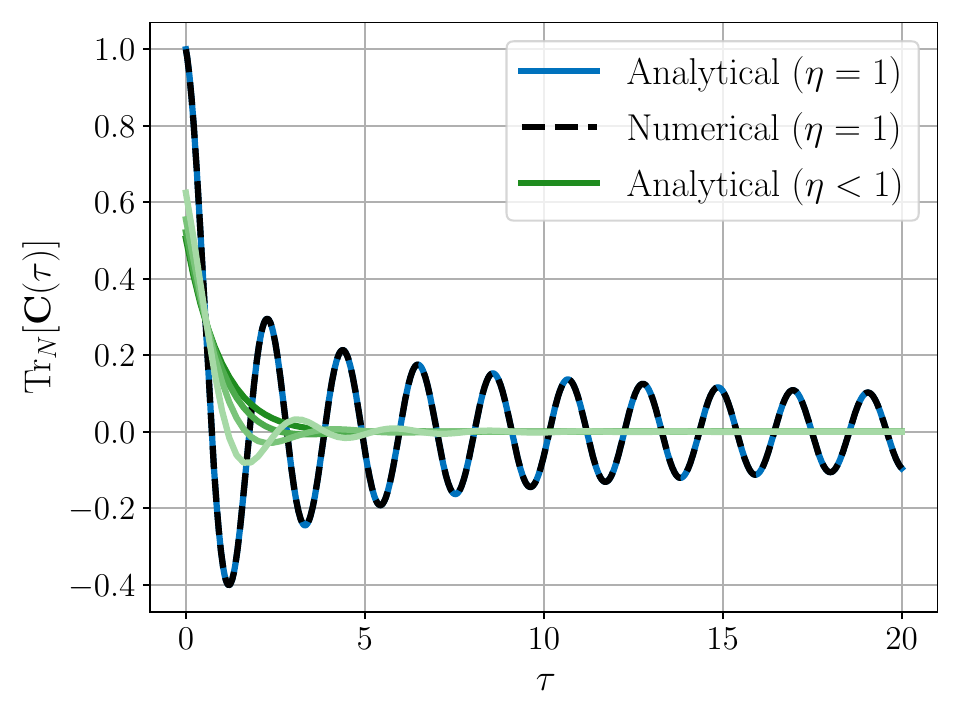}
		\caption{NAAC $\lim_{N\to \infty}\Tr_{N} [\mathbf{C}(\tau)]$ at $\proximity\to1^-$: analytically (continuous, blue) and numerically (dashed, black) along with $\proximity<1$: (green). As one sees, even at $\proximity=0.8$, the decay is exponential and vanishes quickly. However, at $\proximity\to1^-$, the exponential terms cancel out, leaving a slow decaying oscillation, as in Lemma~\ref{lemma:crit-decay}. As clear from the figure, all values of NAAC at $\tau=0$ fall inside the interval $[1/2,1]$, which validates Eq.~\eqref{eq:c-0}.}\label{fig:critical-decay} 
	\end{figure}
	
	\begin{figure}[tb]
		\centering
		\includegraphics[width=0.98\linewidth]{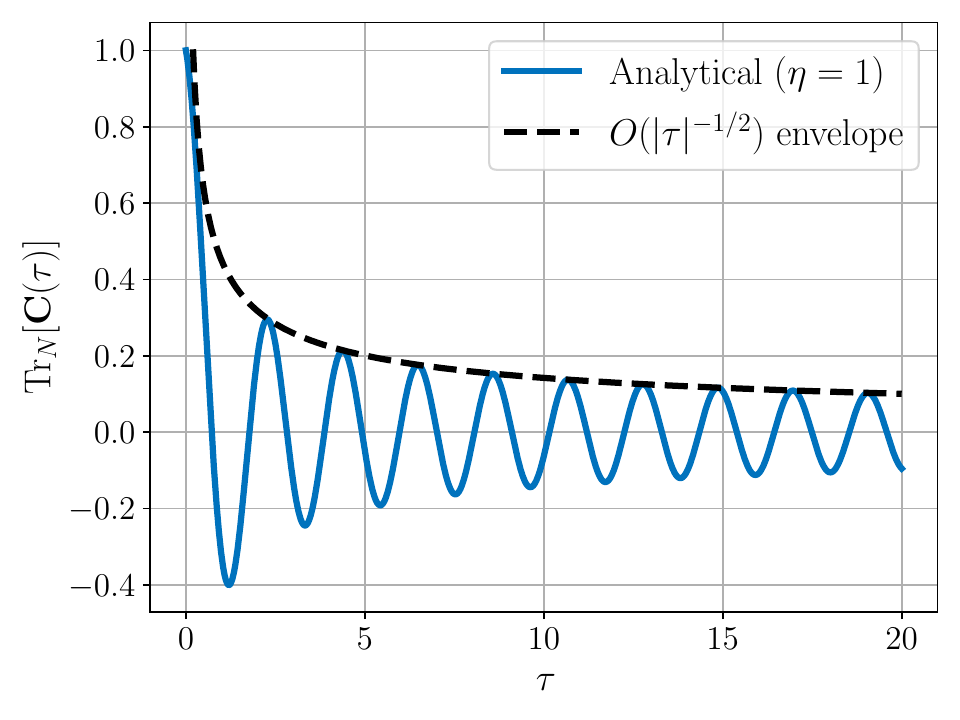}
		\caption{NAAC $\lim_{N\to \infty}\Tr_{N} [\mathbf{C}(\tau)]$ at $\proximity=1$: analytically (solid, blue) and its $O(\tau^{-1/2})$ envelope (dashed, black), presented in Lemma~\ref{lemma:crit-decay}.}\label{fig:slow-decay}
	\end{figure}
	
	Figure~\ref{fig:slow-decay} shows NAAC at the edge of instability as a function of time delay. The solid blue line is the analytically derived NAAC (see Theorem~\ref{thr:naac-crit}) and the dashed black line represents the $O(|\tau|^{-1/2})$ envelope, as evident in the asymptotic expression given in Lemma~\ref{lemma:crit-decay}. Specifically, the black line is the first term of Eq.~\eqref{eq:C-critical-decay}, without the cosine function and the higher order terms.

	\medskip
	
	The  time-domain description has important physical implications. The block-GOE network in the large-$N$ limit behaves like a high-dimensional \emph{underdamped} stochastic oscillator, a fact that is mathematically expressed by the complex exponents of $\pm \iu \proximity \freq \tau$ in Eq.~\eqref{eq:closed-ctau}. These NAAC pseudo-oscillations yield actual harmonic terms for large $|\tau|$ at the edge of instability, as discussed in Lemma~\ref{lemma:crit-decay}. Unlike the stochastic oscillator, however, block-GOE networks show a nontrivial, power-law, slow decay near the edge of instability. This is consistent with the perspective of  pseudo-Hopf bifurcation, observed by Calvo \emph{et al.}, because the oscillations at large $|\tau|$  emerge only at the edge of instability~\cite{calvo_robust_2026}.  
	
	Lemma~\ref{lemma:crit-decay} offers another important insight. Criticality in neural systems is known to be linked to \emph{long-range temporal correlations} (LRTC)~\cite{botcharova_markers_2014}, which represent \emph{``persistence''}, that is, the ability of the brain to ``memorize'', thus passing information  forward in time and enabling learning capabilities. Such LRTC are known so far for disordered, i.i.d. Gaussian connectivity structures~\cite{Hu2022-kf, calvo_robust_2026}. To our knowledge, this is the first time they are observed in oscillatory (block-GOE) networks.
	
	This finding opens new paths for the analysis of clinical data, as brain activity can show both exponential~\cite{murray_hierarchy_2014, gao_neuronal_2020} and subexponential~\cite{hardstone_detrended_2012, linkenkaer-hansen_long-range_2001, miller_power-law_2009, he_temporal_2010} autocovariance functions. For example, existing methods that focus in the frequency domain (specifically in the shape of the tail of the empirical eigenvalue distribution~\cite{PhysRevLett.133.208401, Hu2022-kf}) can be combined with time-domain fits of $\Tr_{N}[\mathbf{C}(\tau)]$. In particular, time-domain analysis can be used to make the estimate of the proximity to instability $\proximity$ more robust in real-world neural time series. In addition, combining covariance-eigenvalue analysis with fits of $\Tr_{N}[\mathbf{C}(\tau)]$ would allow joint tests of the predicted covariance structure and temporal relaxation, providing additional constraints on candidate dynamical models. Diverging correlation times are very common in critical biological systems~\cite{linkenkaer-hansen_long-range_2001, lombardi_long-range_2021}. 
		
	\section{Conclusions}
	
	In summary, we analytically derived exact asymptotic expressions for the network-averaged power spectrum, network-averaged autocovariance function, and normalized effective dimensionality in a block-Gaussian orthogonal ensemble network. We proved that the NAPS singularity emerges only at the characteristic frequency $\omega_C$ and at the edge of instability $\proximity\to 1^{-}$. Near the singularity NAPS scales with a power law of $O\left(|\omega-\omega_C|^{-1/2}\right)$. We also demonstrated that at the edge of instability the normalized effective dimensionality of the network vanishes at the characteristic frequency, a fact that was empirically known but had not been analytically demonstrated. As the system approaches collapse, the NED vanishes according to the scaling law  $O\left(|\omega-\omega_C|^{3/2}\right)$.
	
	The network-averaged autocovariance function (NAAC), which represents the Fourier transform of NAPS, is expressed in terms of the confluent hypergeometric function of the first kind $\Phi_1$, a non-elementary function that is typically defined via the Humbert series. Asymptotically, NAAC displays pseudo-oscillations, which manifest themselves as complex exponential terms, and an exponential decay with respect to time lag $\tau$, for all $\proximity<1$. At the edge of instability, $\proximity\to 1^{-}$, a simplified expression has been derived, which features the modified Bessel functions of the first kind of orders zero and one. In addition, for sufficiently large $|\tau|$, NAAC displays frequency oscillations at $\omega_C=\freq=\beta/\alpha$ combined with slow decay, which is  inversely proportional to the square root of the time lag.
	
	These results provide new insight into random neural networks. In i.i.d. Gaussian networks, NAPS  follows the power law $1/\omega^2$ at the edge of instability.  NAPS in block-GOE networks diverges asymptotically as $|\omega-\omega_C|^{-1/2}$, which is a slower decay than i.i.d. Gaussian networks. In addition, in the latter case the network-averaged autocovariance function diverges to infinity as $\proximity\to 1^{-}$. Hence, NAPS does not correspond to the Fourier transform of NAAC for such networks at the edge of instability. In contrast,  we have proved that block-GOE networks preserve Fourier-transform duality at the edge of instability.
	
	We have also shown that for block-GOE networks at the edge of instability the normalized effective dimensionality vanishes as $|\omega-\omega_C|^{3/2}$ as the characteristic frequency is approached. In contrast, i.i.d. Gaussian networks display a faster, quartic rate of convergence, that is, $\omega^4$, near the origin.

	In conclusion, our study shows that the limiting behavior (divergence or  collapse to zero) of network observables at a given frequency $\omega_C$ does not fully describe the network behavior, because the latter also involves the rates of convergence (divergence). Due to the fact that block-GOE networks converge to instability at a slower rate than i.i.d. Gaussian networks, they display rich features at the edge of instability, such as the slow decay of NAAC.

	\begin{acknowledgments}
		We thank Arif Babul (University of Victoria, Canada) and Naznin Virjil-Babul (University of British Columbia, Canada)  for helpful discussions and their interest in this work. We are also grateful to Ruben Calvo Ibáñez for kindly clarifying several points regarding Ref.~\cite{PhysRevLett.133.208401} in correspondence.
	\end{acknowledgments}
	
	\appendix

	\section{Notation of the eigenvalue distribution moments, $\mathbb{E}[\lambda^{n}(\omega)]$}
	\label{app:notation}
	
	To derive the expression of $\mathbb{E}[\lambda(\omega)]$ in the large-$N$ limit, where the empirical eigenvalue distribution of $\mathbf{S}(\omega)$ converges to the limiting spectral distribution.
	
	A random eigenvalue  $\mu\in\mathbb{R}$ of the matrix $g\Jmatrix$ in the large-$N$ limit generates two branches of SDM eigenvalues $\mathbf{S}(\omega)$ corresponding to $\sigma=\pm 1$ as follows
	\begin{align}
		& h(\mu; \omega, \alpha,\beta,\s) = \frac{1}{(1-\alpha \mu)^2 + (\omega + \sigma\beta\mu)^2}\,. \label{eq:h}
		\intertext{The above holds for any $\mu$,  given $\alpha, \beta \in\mathbb{R} \backslash \{0\}$, and $\sigma \in\{-1,1\}$~\cite{PhysRevLett.133.208401}. Since $\lambda(\omega)$ is connected to the eigenvalues of $g\Jmatrix$ via Eq.~\eqref{eq:h}, the expectation of the random eigenvalue $\lambda(\omega)$ of the SDM $\mathbf{S}(\omega)$ is given by}
		&\mathbb{E}[\lambda(\omega)] = \mathbb{E}_{\mu\sim f_{g\Jmatrix}}[h(\mu;\omega, \alpha,\beta,\s)] \notag\\
		&\quad= \frac{1}{2} \sum_{\sigma\in\{-1,1\}} \int_{-2g}^{2g} f_{g\Jmatrix}(\mu;g) h(\mu;\omega, \alpha,\beta,\s)\, \dd\mu, \label{eq:h-integral}
	\end{align}
	where $f_{g\Jmatrix}(\mu;g)$ is the probability density function of the Wigner's semicircle law for the GOE matrix $g\Jmatrix$, evaluated at eigenvalue $\mu\in[-2g, 2g].$ Respectively, $\mathbb{E}_{\mu\sim f_{g\Jmatrix}}[\cdot]$ denotes the expectation over the probability distribution of $g\Jmatrix$. More generally, the $n$-th moment of the limiting eigenvalue distribution of $\mathbf{S}(\omega)$ is given by
	\begin{align}
		&\mathbb{E}[\lambda^{n}(\omega)] \notag \\
		&\quad= \mathbb{E}_{\mu\sim f_{g\Jmatrix}}[h^n(\mu;\omega, \alpha,\beta,\s)] \notag\\
		&\quad= \frac{1}{2}\sum_{\sigma\in\{-1,1\}} \int_{-2g}^{2g} f_{g\Jmatrix}(\mu;g) h^{n}(\mu;\omega, \alpha,\beta,\s)\, \dd\mu, 
		\label{eq:h-n-integral}
	\end{align}
	for all $\omega\in\mathbb{R}$, $n\in\{1,2\}$. Although Eq.~\eqref{eq:h-n-integral} holds for all $n \in \mathbb{N},$ as well as for $n=0$, trivially, we focus only on the two first statistical moment of $\lambda(\omega)$. We substitute Eq.~\eqref{eq:h} into Eq.~\eqref{eq:h-n-integral} to obtain
	\begin{align*}
		&\mathbb{E}[\lambda^{n}(\omega)] = \frac{1}{4 \pi g^2}\sum_{\sigma} \int_{-2g}^{2g}
		\frac{\sqrt{4g^2-\mu^2}}{\left[ (1-\alpha \mu)^2 + (\omega + \s\beta\mu)^2\right]^{n}}\, \dd\mu,
	\end{align*}
	for all $\alpha, \beta \in \rr$, $g \in [0,g_C)$, $n \in \{1,2\}$. Using the change of variables $\mu/g \mapsto u$, the above integral transforms into
	\begin{align}
		&\mathbb{E}[\lambda^{n}(\omega)]= \frac{1}{4 \pi }\sum_{\sigma} 
		\int_{-2}^{2}\frac{\sqrt{4-u^2}}{\left[ (1-\alpha g u)^2 + (\omega + \s\beta g u)^2\right]^{n}}\, \dd u, 
	\end{align}
	for all $\omega \in\mathbb{R}$.
	
	To remove parametric redundancy, as in Eq.~\eqref{eq:new-W}, we use the transformations 
	\(2\alpha g \mapsto \proximity\) and \(\beta/\alpha \mapsto \freq\), which leads to

	\begin{align}
		&\mathbb{E}[\lambda^{n}(\omega)]   =  
		\frac{1}{4 \pi }\sum_{\sigma\in\{-1,1\}} \notag \\
		&\qquad\times\int_{-2}^{2}\frac{\sqrt{4-u^2}}{\left[ (1- \proximity u / 2)^2 + (\omega + \s\freq \proximity u /2)^2 \right]^{n}}\, \dd u,\label{eq:h-integ-n-eta-rho}
		\intertext{for all $\omega\in\rr$ and proximity to instability $\proximity\in [0,1)$, $\freq \in \mathbb{R}$, $n \in \{1, 2\}$. From Eq.~\eqref{eq:h-integ-n-eta-rho}, we can see that if we apply the transformation $\freq \to -\freq$, it is equivalent to flipping both signs $\s \to -\s$. Due to this property, both statistical moments of $\lambda(\omega)$, $\ee[\lambda^n(\omega)]$, $n=1,2$, are invariant to the sign of $\freq$. Hence, without loss of generality, $\freq>0$, and we study $\freq=0$ separately in Appendix~\ref{app:rho-0}. In addition, in all above equations the symmetry $\omega\to-\omega$ is also visible, which is expected as $\lambda(\omega)$ is a random SDM eigenvalue. We set the parameter vector $\abxi\triangleq (\proximity, \freq, \s)^{\top}$, such that $\abxi \in \boldsymbol{\Theta}$, where $\boldsymbol{\Theta}=[0,1)\times \rr_+\times \{ -1,1\}$. We complete the square in Eq.~\eqref{eq:h-integ-n-eta-rho} to obtain}
		&\quad= \frac{(1 + \freq^2)^n}{4 \pi }\sum_{\sigma\in\{-1,1\}} 
		\notag \\ 
		&\qquad \times
		\int_{-2}^{2}\frac{\sqrt{4-u^2}}{\left[A^2(\omega;\freq,\s) + [\C u + B(\omega;\freq,\s)]^2  \right]^{n}}\, \dd u,\label{eq:n-mom-a-b-c}
	\end{align}
	for all $\omega\in\rr$ where 
	\begin{align}
		&A(\omega;\freq,\s) = |\freq+\s \omega| \notag \\
		&B(\omega;\freq,\s) = \s\freq\omega - 1 \notag \\
		&\C = \frac{1}{2} \proximity (1 + \freq^2).
	\end{align}

	From this point on, we will be working with the parameters $\proximity, \freq$ instead of $\alpha, \beta, g$. Consider the functions $\phi, \psi$, defined as
	\begin{align}
		\phi(u) &= \sqrt{4 - u^2}  \notag \\ 
		\psi(u; \omega, \abxi) &= \frac{1}{A^2(\omega;\freq,\s) + [\C u + B(\omega;\freq,\s)]^2}\,, \label{eq:phi-psi}
	\end{align}
	where $\operatorname{dom} \phi = [-2,2]$,  $\operatorname{dom} \psi = \rr$, $\abxi \in \boldsymbol{\Theta}$. Both functions in Eq.~\eqref{eq:phi-psi} are real-valued, where $\psi$ is a \emph{Lorentzian} function, translated by $B(\omega;\freq,\s)$, and $\phi$ is a semi-circle of radius 2. Define 
	\begin{align}
		\Lambda_{n}(\omega;\abxi) &\triangleq \frac{(1+\freq^2)^n}{2\pi}\int_{-2}^{2} \phi(u)\psi^{n}(u; \abxi) \, \dd u, \label{eq:Lambda-def}
	\end{align}
	for all $n \in \{1,2\}$. Notice that, by definition,
	\begin{align}
		\ee[\lambda^n(\omega)] = \frac{1}{2} \sum_{\s \in \sdom} \Lambda_{n}(\omega;\abxi), \label{eq:A10}
	\end{align}
	for all $\omega \in \mathbb{R}$. In the subsequent appendices, we derive exact formulae of $\ee[\lambda^n(\omega)]$, for $n=1,2$.
	
	\section{First moment of the eigenvalue distribution, $\mathbb{E}[\lambda(\omega)]$}
	\label{app:naps}
	\subsection{Derivation}
	For the computation of the $\Lambda_1(\omega;\abxi)$ function, for all $\omega\in\rr$, we will make the following arguments. We start from Eq.~\eqref{eq:Lambda-def}, and notice that $\phi$ is a real-valued function, $\phi(u)=\overline{\phi(u)}$, for all $u \in [-2,2]$, where $\overline{\phi(u)}$ denoted the complex conjugate of $\phi(u)$. This way, we can apply the Plancherel-Parseval identity~\cite[see][Sec.~2.2.2, Thr.~2.2.14]{grafakos_classical_2014}, to obtain
	\begin{align}
		\Lambda_{1}(\omega;\abxi) &= \frac{1+\freq^2}{2\pi}\int_{-\infty}^{\infty} \mathcal{F}\phi(x) \mathcal{F}\psi(x; \omega, \abxi) \, \dd x, \label{eq:Lambda-fourier-integ}
	\end{align}
	for all $\omega \in \rr$, where $\mathcal{F}[\cdot]$ denotes the Fourier transform with the exponent of $-\iu 2 \pi x u$~\cite[see][Sec.~2.2.2, Def.~2.2.8]{grafakos_classical_2014}. Since $\phi$ is an a real, even function, $\mathcal{F}\phi$ is also a real even function, so it is trivially equal to its complex conjugate, for all $u\in \rr$. We cite the specific equation of $\mathcal{F}\phi$ in the following paragraph.
	
	The Fourier transform of $\phi$ is given by~\cite[see][Sec.~1.3, Eq.~(8)]{Erdelyi1954TablesVol1}
	\begin{align}
		& \mathcal{F}\phi (x)=\frac{1}{x} J_1(4\pi x), \label{eq:fourier-semicirc}
	\end{align}
	for all $x \in \rr \setminus \{0\}$, where $J_1$ denotes the \emph{Bessel function of the first kind} and of order 1~\cite[see][Ch.~III, Sec.~3.1]{Watson1995Bessel}.
	
	The Fourier transform of $\psi$ is given by~\cite{campbell_fourier_1948}
	\begin{align}
		&\mathcal{F}\psi(x;\omega,\abxi) = \frac{\pi }{A(\omega;\freq,\s) \C}
		\notag \\
		&\quad \times \exp \left(\frac{2\pi}{\C}\left[-A(\omega;\freq,\s) |x| +\iu B(\omega;\freq,\s) x \right]\right), \label{eq:fourier-lorentz}
	\end{align}
	for all $x \in \rr$, and for all $\omega \in \rr \setminus \{-\freq\s\}$, in order to exclude the zero of $A(\omega;\freq,\s)$.
	
	By substituting Eq.~\eqref{eq:fourier-semicirc} and Eq.~\eqref{eq:fourier-lorentz} into the Fourier integral of Eq.~\eqref{eq:Lambda-fourier-integ}, we can express $\Lambda_1(\omega;\abxi)$ as follows:
	\begin{align}
		&\Lambda_1(\omega;\abxi) = \frac{2\pi}{\proximity A(\omega;\freq,\s)} \notag \\ 
		&\quad \times \int_{-\infty}^{\infty} \frac{J_1(4\pi x)}{x} e^{{2\pi}\left[-A(\omega;\freq,\s) |x| +\iu B(\omega;\freq,\s) x \right]/{\C}} \, \dd x. \label{eq:lambda-integ}
		\intertext{Eq.~\eqref{eq:lambda-integ} resembles the Laplace transform of the function $x^{-1}J_1(4\pi x)$ over the real numbers. Using~\cite[see][Sec.~{4.14}, Eq.~(5)]{Erdelyi1954TablesVol1},}
		&\Lambda_1(\omega;\abxi) =\frac{\sqrt{2}}{\proximity^2(1+\freq^2)A(\omega;\freq,\s)} \notag \\ 
		&\quad \times \Bigg[ \sqrt{H^2(\omega;\abxi) + 4A^2(\omega;\freq,\s)B^2(\omega;\freq,\s)} \notag \\
		&\qquad - H(\omega;\abxi)
		\Bigg]^{1/2} - \frac{2}{\proximity^2(1+\freq^2)},\label{eq:lambda1-closed}
	\end{align}
	for all $\omega \in \rr \setminus \{-\s \freq \}$, where 
	\begin{align}
		H(\omega;\abxi)=B^2(\omega;\freq,\s)-A^2(\omega;\freq,\s)-4\C^2.
	\end{align}
	Despite the fact that $A(\omega;\abxi)=0$ for $\omega=-\s\freq$, the limit of $\Lambda_{1}(\omega;\abxi)$ is finite for all $\proximity \in (0,1)$ (see more in Appendix~\ref{app:exact-om-rho}), and is unbounded only for $\proximity\to1^-$ and $\omega\to-\s\freq$. The hole in $-\s\freq$ is an artifact of the Plancherel-Parseval argument (see Eq.~\eqref{eq:Lambda-fourier-integ}), and has no physical meaning for the system. In fact, it is possible to substitute in Eq.~\eqref{eq:phi-psi} for $\omega=-\freq \s$ and analytically calculate the integral of Eq.~\eqref{eq:Lambda-def}, for all $\omega \in \rr$, provided that $\proximity \in (0,1)$.
	
	\subsection{Behavior close to $\omega_C$}\label{app:behavior-close-to-char}

	The goal of this appendix is three fold; (1) to prove that $\ee[\lambda(\omega)]$ displays an infinity only at $\proximity \to 1^-$ and $\omega \to \pm \freq$, (2) to investigate the behavior of $\ee[\lambda(\omega)]$ near $\pm \freq$ at the edge of instability, and (3) to extend this analysis beyond exactly $\proximity\to 1^-$. In one sentence, this appendix aims to analyze how $\ee[\lambda(\omega)]$ behaves $\proximity$-close to instability and $\varepsilon$-close to $\omega_C$.
	
	Before starting the analysis, we define the function $\mathcal{K}(\omega;\abxi)$, calculated at $\omega \in \rr$, as
	\begin{align}
		&\mathcal{K}(\omega;\abxi) \notag \\ &\quad\triangleq \sqrt{H^2(\omega;\abxi) + 4A^2(\omega;\freq,\s) B^2(\omega;\freq,\s)} -H(\omega;\abxi), \label{eq:k}
	\end{align}
	for all $\omega \in \rr$. Eq.~\eqref{eq:k} is linked to $\Lambda_1(\omega;\abxi)$, for all $\omega \in \rr \setminus \{ -\s \freq\}$ via the relation
	\begin{align}
		\Lambda_1(\omega;\abxi) = \frac{\sqrt{2 \mathcal{K}(\omega;\abxi)}}{\proximity^2(1 + \freq^2)A(\omega;\freq,\s)} - \frac{2}{\proximity^2(1+\freq^2)}. \label{eq:lambda-k}
	\end{align}
	
	\subsubsection{Behavior at $\omega=\freq $}\label{app:exact-om-rho}
	We are interested in the limits of $\mathcal{K}$ near the point of $A(\omega;\freq,\s)=0$. This value is obtained if and only if $\omega=-\s \freq$. We substitute the frequency $-\s\freq$ to obtain
	\begin{align}
		H(-\s\freq;\abxi) = (\freq^2+1)^2(1-\proximity^2), \label{eq:h-s-r}
	\end{align}
	for all $\proximity \in (0,1)$, $\freq>0$. For any value of $\proximity\in(0,1)$, Eq.~\eqref{eq:h-s-r} is positive, and, due to the fact that $H$ is a continuous function, $H$ is positive \emph{near} $-\s\freq$. Since $H(\omega;\abxi)>0$, for all $\omega$ sufficiently close to $-\s\freq$, we can simplify $\mathcal{K}(\omega;\abxi)$, for all $\omega $ in this region around $-\s\freq$, to obtain
	\begin{align}
		&\frac{\mathcal{K}(\omega;\abxi)}{A^2(\omega;\freq,\s)} = \frac{H(\omega;\abxi)}{A^2(\omega;\freq,\s)} 
		\notag \\ 
		&\quad \times \left[\sqrt{1+\frac{4A^2(\omega;\freq,\s)B^2(\omega;\freq,\s)}{H^2(\omega;\abxi)}}-1\right], \label{eq:k-over-a}
		\intertext{for all $\omega$ sufficiently close to $-\s\freq$. As $\omega$ approaches $-\s\freq$, the denominator of $A^2(\omega;\abxi)$ approaches zero, while, at the same time, the term in the brackets also approaches zero. To deal with this, we rationalize the numerator, by multiplying both the numerator and the denominator by the radical conjugate. This yields to}
		&\lim_{\omega \to -\s \freq} \frac{\mathcal{K}(\omega;\abxi)}{A^2(\omega;\freq,\s)}=\frac{2}{1-\proximity^2}. \label{eq:lim-s-r}
		\intertext{We substitute Eq.~\eqref{eq:lim-s-r} into Eq.~\eqref{eq:lambda-k} to obtain}
		&\lim_{\omega \to -\s \freq} \Lambda_1(\omega;\abxi)=\frac{2}{(1+\freq^2)\proximity^2\sqrt{1-\proximity^2}} -\frac{2}{\proximity^2(1+\freq^2)}. \label{eq:lim-lambda-neg-sig-rho}
	\end{align}
	As one would expect, the limit in Eq.~\eqref{eq:lim-lambda-neg-sig-rho} does not depend on the specific sign of $\s$, only at the sign of $\omega \s$ (keeping a fixed $\freq>0$). The limit of $\ee[\lambda(\omega)]$ at $\omega \to \pm \freq$ is connected to Eq.~\eqref{eq:lim-lambda-neg-sig-rho} via
	\begin{equation}
		\lim_{\omega \to \pm \freq} \ee[\lambda(\omega)] = \frac{1}{2}\bigg[\lim_{\omega \to -\s \freq} \Lambda_1(\omega;\abxi)+\lim_{\omega \to \s \freq} \Lambda_1(\omega;\abxi)\bigg], \label{eq:ee-lim-s-r}
	\end{equation}
	for all  $\omega \in \rr$. The limit of $\Lambda(\omega;\abxi)$ for $\omega \to \s \freq$ is finite for all $\proximity \in (0,1)$, as well as for $\proximity\to 1^{-}.$ Eq.~\eqref{eq:ee-lim-s-r} provides another intuitive validation that $\ee[\lambda(\omega)]$ is an even function of $\omega \in \rr$; when $\omega>0$, the singularity applies at $-\freq$, and respectively for $\omega<0$ at $\freq$. Clearly, based on Eq.~\eqref{eq:ee-lim-s-r} and Eq.~\eqref{eq:lim-lambda-neg-sig-rho}, $\ee[\lambda(\omega)]$ tends to infinity at $\omega\to\pm\freq$, for $\proximity\to 1^-$. In the following appendix, we analyze its behavior arbitrarily close to $\pm\freq$.
	
	\subsubsection{Behavior at $\omega=\freq+\varepsilon$ and $\proximity\to 1^-$}\label{app:B2b}
	
	As discussed in the previous appendix, the singularity of $\ee[\lambda(\omega)]$ at $\omega \to \pm \freq$ comes from the singularity of $\Lambda_1(\omega;\abxi)$ as $\omega\to - \s\freq $. Here, we analyze $\Lambda_1(\omega;\abxi)$ at a neighborhood near $-\s \freq $, at $\proximity\to 1^-$. Specifically, we introduce a small perturbative variable $\varepsilon \in \rr$ and work in frequency $\omega =- \s \freq +\varepsilon$. Substituting yields
	\begin{align}
		& A(-\freq\s+\varepsilon;\freq,\s) = |\varepsilon|, \notag
		\\
		& B(-\freq\s+\varepsilon;\freq,\s) = -(\freq^2+1) + \s \freq \varepsilon,\notag
	\end{align}
	for an arbitrary $\varepsilon \in \rr.$ At the edge of instability, namely $\proximity\to 1^-$, 
	\begin{align*}
		& \lim_{\proximity \to 1^-}H(-\s\freq + \varepsilon) = \varepsilon\left[ 
		(\freq^2-1)\varepsilon - 2\s\freq(\freq^2+1)
		\right],
	\end{align*}
	for all $\varepsilon \in \rr$. Substituting into Eq.~\eqref{eq:k-over-a} yields
	\begin{align}
		&\frac{\mathcal{K}(\omega;\abxi)}{A^2(\omega;\freq,\s)}=\frac{1}{|\varepsilon|}|(\freq^2 - 1)\varepsilon - 2\freq\s(\freq^2 + 1)| \notag \\
		&\quad \times \Bigg[\sqrt{1 + \frac{4[-(\freq^2+1) + \s \freq \varepsilon]^2}{[(\freq^2 - 1)\varepsilon - 2\s\freq(\freq^2+1)]^2} } \notag \\
		&\qquad - \sgn\left(  \varepsilon\left[ 
		(\freq^2-1)\varepsilon - 2\s\freq(\freq^2+1)
		\right]\right) \Bigg]. \label{eq:ka-prefinal-frac}
	\end{align}
	The bracketed term approaches a bounded, nonzero term as $\varepsilon\to 0$. At the same time, the first term approaches infinity with a dominant order of $O(1/\varepsilon).$ Therefore, at $\proximity\to1^-$,
	\begin{equation}
		\frac{\mathcal{K}(\omega;\abxi)}{A^2(\omega;\freq,\s)}=\frac{1}{|\varepsilon|}
		\left[
		2(\freq^2+1)\left(\sqrt{1+\freq^2} + \s \freq \, \sgn(\varepsilon)\right).
		\right]. \label{eq:ka-final-frac}
	\end{equation}
	As $\proximity\to 1^-$, by substituting Eq.~\eqref{eq:ka-final-frac} into Eq.~\eqref{eq:lambda1-closed} yields
	\begin{equation}
		\ee[\lambda(\omega)] \sim \scale_{\pm}(\freq) |\omega-\omega_C|^{-1/2} , \label{eq:crit-asymp}
	\end{equation}
	for all $\omega$ sufficiently close to $\omega_C=\pm\freq$, where for given $\proximity\to1^-$ and $ \freq>0$, $q(\freq)$ evaluates from taking the limit of Eq.~\eqref{eq:ka-final-frac} for sufficiently small $\varepsilon$, as 
	\begin{align}
		\scale_{\pm}(\freq)&=\frac{\left(\sqrt{1+\freq^2} \mp \freq \right)^{1/2}}{\sqrt{1+\freq^2}} \notag \\ 
		&= \left( 1 + \freq^2\right)^{-1/4} \left| \cos\left(\frac{\phi}{2}\right) \mp \sin \left(\frac{\phi}{2}  \right)\right|,\label{eq:crit-asymp-scale}
	\end{align}
	where $\phi=\arctan \freq$. In Eq.~\eqref{eq:crit-asymp-scale}, the positive sign corresponds to $\omega>\omega_C$ and the negative to $\omega<\omega_C$. More compactly, 
	\begin{equation}
		\ee[\lambda(\omega)] = \order{|\omega-\omega_C|^{-1/2}}, \label{eq:crit-asymp-o}
	\end{equation}
	for all $\omega$ sufficiently close to $\omega_C=\pm\freq$.
	
	\subsubsection{Behavior at $\omega=\proximity\freq +\varepsilon$}\label{app:singularity-asympt-subcrit}
	
	In this appendix, we derive the asymptotic behavior of $\mathbb{E}[\lambda(\omega)]$
	near the characteristic frequency $\omega_C = \eta\kappa$ in the subcritical regime,
	that is, for $\eta < 1$. Since $\mathbb{E}[\lambda(\omega)]$ is an even function of
	$\omega$ (see Appendix~\ref{app:notation}), we assume $\omega > 0$ without loss of generality.
	
	In Appendix~\ref{app:B2b} we showed that, at $\eta \to 1^-$, $\mathbb{E}[\lambda(\omega)]$ diverges as $|\omega - \omega_C|^{-1/2}$, with amplitudes $q_{\pm}(\kappa)$ that differ on each side of the singularity (see Eqs.~\eqref{eq:crit-asymp} and \eqref{eq:crit-asymp-scale}). A natural question is whether this law persists for $\eta < 1$. For any fixed $\eta \in (0,1)$ the answer is negative. The limit in Eq.~\eqref{eq:lim-lambda-neg-sig-rho} is finite, and
	$\mathbb{E}[\lambda(\omega)]$ displays a smooth, bounded peak in the neighborhood of
	$\omega_C$. The divergence derived in Appendix~\ref{app:B2b} therefore results from two
	limits, $\omega \to \omega_C$ and $\eta \to 1^-$, which do not commute. To describe how
	the bounded peak turns into the singularity of Eq.~\eqref{eq:crit-asymp}, we let
	$\varepsilon \triangleq \omega - \eta\kappa$ and $1 - \eta$ tend to zero together.
	
	The relevant scales follow from the structure of Eq.~\eqref{eq:n-mom-a-b-c}. For fixed
	$\omega \in \mathbb{R}$, the integrand is the product of the semicircle $\phi(u)$ and the
	Lorentzian $\psi(u;\omega,\boldsymbol{\theta})$. The Lorentzian is centered at
	$u_0 = -B(\omega;\kappa,\sigma)/C_{\eta,\kappa}$ and has half-width
	$A(\omega;\kappa,\sigma)/C_{\eta,\kappa}$ (see Eq.~\eqref{eq:phi-psi}). For $\sigma = +1$, the
	half-width $A(\omega;\kappa,1)/C_{\eta,\kappa}$ is of order one, and the corresponding
	term in Eq.~\eqref{eq:A10} is bounded near $\omega_C$. For $\sigma = -1$, we substitute
	$\omega = \eta\kappa + \varepsilon$ to obtain
	\begin{align}
		u_0 = 2 + \frac{2\left[(1-\eta) + \kappa\varepsilon\right]}{\eta(1+\kappa^2)},
		\label{eq:B18a}
	\end{align}
	and
	\begin{align}
		\frac{A(\omega;\kappa,-1)}{C_{\eta,\kappa}}
		= \frac{2\left|(1-\eta)\kappa - \varepsilon\right|}{\eta(1+\kappa^2)}, \label{eq:B18b}
	\end{align}
	for all $\varepsilon \in \mathbb{R}$.
	
	At $\eta \to 1^{-}$ and $\varepsilon = 0$, the Lorentzian collapses onto the edge $u = 2$ of the support, where $\phi$ vanishes as a square root. This collapse is the origin of the singularity in Appendix~\ref{app:B2b}. Equations~\eqref{eq:B18a} and~\eqref{eq:B18b} show that, for $\eta < 1$, the offset of the Lorentzian from the edge and its width are both of order $1-\eta$. A shift in frequency domain by $\varepsilon$ displaces the center by an amount of order $\varepsilon$. The method of dominant balance \cite{barenblatt1996scaling, bender1999advanced} therefore distinguishes three regimes.
	
	If $|\varepsilon| \ll 1-\eta$, the shift is negligible compared with the offset and the
	width. The overlap of $\phi$ and $\psi$ is then insensitive to $\varepsilon$, and
	$\mathbb{E}[\lambda(\omega)]$ is locally flat. This is the inner region. In it, a regular
	expansion in powers of $\varepsilon$ holds, with coefficients that diverge as
	$\eta \to 1^-$, as in Appendix~\ref{app:rho-0-close-omega-0} for $\kappa = 0$.
	
	If $|\varepsilon| \gg 1-\eta$, the terms of order $1-\eta$ in Eq.~\eqref{eq:B18a} and Eq.~\eqref{eq:B18b} are negligible, and the analysis of Appendix~\ref{app:B2b} applies. This is the outer region.
	
	The distinguished limit, in which all terms in Eq.~\eqref{eq:B18a} and Eq.~\eqref{eq:B18b} are of the same order, is
	\begin{equation}
		\varepsilon = (1-\eta)X,
		\label{eq:B19}
	\end{equation}
	where $X$ is an $\order{1}$ term.
	
	In the distinguished limit, the semicircle contributes to the integral at leading order
	only through its edge layer, $2 - u = \order{1-\eta}$. Accordingly, in
	Eq.~\eqref{eq:n-mom-a-b-c} with $\sigma = -1$ we rescale the integration variable as
	\begin{equation}
		u = 2 - 2(1-\eta)y,
		\label{eq:B20}
	\end{equation}
	where $y \in [0, 2/(1-\eta)]$. Substituting Eq.~\eqref{eq:B19} and Eq.~\eqref{eq:B20}
	yields the exact relations $1 - \eta u/2 = (1-\eta)(1+\eta y)$,
	$\omega - \kappa\eta u/2 = (1-\eta)(X + \kappa\eta y)$, and
	$4 - u^2 = 8(1-\eta)y\left[1 - (1-\eta)y/2\right]$. Hence, the term $\sigma = -1$ of
	Eq.~\eqref{eq:n-mom-a-b-c} for $n = 1$ becomes
	\begin{equation}
		\frac{\sqrt{2}}{\pi}(1-\eta)^{-1/2}
		\int_0^{2/(1-\eta)}
		\frac{\sqrt{y}\left[1 - (1-\eta)y/2\right]^{1/2}}
		{(1+\eta y)^2 + (X+\kappa\eta y)^2}\, dy,
		\label{eq:B21}
	\end{equation}
	for all $X \in \mathbb{R}$.
	
	As $\eta \to 1^-$ with $X$ fixed, the integrand converges to
	$\sqrt{y}/\left[(1+y)^2 + (X+\kappa y)^2\right]$, and the upper limit tends to infinity.
	The limiting integrand decays as $y^{-3/2}$, so the limiting integral converges. Both the
	truncation and the substitution of $\eta$ by $1$ introduce corrections of relative order
	$(1-\eta)^{1/2}$. The bounded term $\sigma = +1$ (see Eq.~\eqref{eq:ee-lim-s-r}) is of the same relative order. Therefore,
	\begin{equation}
		\mathbb{E}[\lambda(\omega)]
		= \frac{\sqrt{2}}{\pi}(1-\eta)^{-1/2}\, \Xi(X)
		\left[1 + {O}\!\left((1-\eta)^{1/2}\right)\right],
		\label{eq:B22}
	\end{equation}
	where
	\begin{equation}
		\Xi(X) \triangleq \int_0^{\infty}
		\frac{\sqrt{y}\, dy}{(1+y)^2 + (X+\kappa y)^2},
		\label{eq:B22b}
	\end{equation}
	for all $X \in \mathbb{R}$ and $\kappa > 0$, at $\eta \to 1^-$.
	
	The integral in Eq.~\eqref{eq:B22b} admits a closed form. Its denominator is the
	quadratic $(1+\kappa^2)y^2 + 2(1+\kappa X)y + 1 + X^2$. Its discriminant is nonpositive,
	since $(1+\kappa X)^2 \leq (1+\kappa^2)(1+X^2)$ by the Cauchy--Schwarz inequality. We
	define $c(X) \triangleq \left[(1+X^2)/(1+\kappa^2)\right]^{1/2}$ and
	$\chi(X) \in [0,\pi)$ through
	\begin{equation}
		\cos\chi(X) = \frac{1+\kappa X}{\left[(1+\kappa^2)(1+X^2)\right]^{1/2}},
	\end{equation}
	for all $X \in \mathbb{R}$. The change of variables $y = c(X)x$ yields
	\begin{equation}
		\Xi(X) = \frac{1}{(1+\kappa^2)\, c^{1/2}(X)}
		\int_0^{\infty} \frac{x^{1/2}\, dx}{x^2 + 2x\cos\chi(X) + 1}.
		\label{eq:B23}
	\end{equation}
	The integral in Eq.~\eqref{eq:B23} is the case $\mu = 3/2$ of
	\begin{equation}
		\int_0^{\infty} \frac{x^{\mu-1}\, dx}{x^2 + 2x\cos\chi + 1}
		= \frac{\pi\sin\left[(1-\mu)\chi\right]}{\sin\chi\,\sin\mu\pi},
	\end{equation}
	which holds for $0 < \mu < 2$ and $0 < \chi < \pi$, with $\chi = 0$ understood as a limit
	\cite{gradshteyn2007table}. For $\mu = 3/2$ it equals
	$\pi/\left[2\cos(\chi/2)\right]$. Using
	$\cos(\chi/2) = \left[(1+\cos\chi)/2\right]^{1/2}$, we obtain
	\begin{equation}
		\Xi(X) = \frac{\pi}{\left[2(1+\kappa^2)\,\Upsilon(X)\right]^{1/2}},
		\label{eq:B24}
	\end{equation}
	where $\Upsilon(X)$ is defined as
	\begin{equation*}
		\Upsilon(X) \triangleq \left[(1+\kappa^2)(1+X^2)\right]^{1/2} + 1 + \kappa X,
	\end{equation*}
	where $\Upsilon(X) > 0$ for all $X \in \mathbb{R}$. Substituting Eq.~\eqref{eq:B24} into
	Eq.~\eqref{eq:B22} yields
	\begin{equation}
		\mathbb{E}[\lambda(\omega)] \sim
		\frac{1}{\left[(1+\kappa^2)(1-\eta)\,\Upsilon(X)\right]^{1/2}},
		\qquad
		X = \frac{\omega - \eta\kappa}{1-\eta},
		\label{eq:B25}
	\end{equation}
	for all $X = \order{1}$, at $\eta \to 1^-$.
	
	Equation~\eqref{eq:B25} describes the full crossover between the smooth peak and the
	singularity. We first consider the peak. The function $\Upsilon$ attains its minimum,
	$\Upsilon(-\kappa) = 2$, at $X = -\kappa$. Therefore,
	\begin{align}
		\max_{\omega}\, \mathbb{E}[\lambda(\omega)]
		&= \left[2(1+\kappa^2)(1-\eta)\right]^{-1/2} \notag \\ 
		&\quad\times \left[1 + {O}\!\left((1-\eta)^{1/2}\right)\right],
		\label{eq:B26}
	\end{align}
	attained at $\omega_* = (2\eta - 1)\kappa + {O}\!\left((1-\eta)^{3/2}\right)$. The
	peak height diverges as $(1-\eta)^{-1/2}$. The peak itself is displaced from
	$\omega_C = \eta\kappa$ toward the origin by $\kappa(1-\eta)$. This displacement reflects
	the square-root vanishing of $\phi$ at the edge of the support. According to
	Eq.~\eqref{eq:B18a} and Eq.~\eqref{eq:B18b}, decreasing $\varepsilon$ moves the center of the Lorentzian into the support, where the semicircle density is larger, at the cost of a wider Lorentzian. The balance of these two effects occurs at $X = -\kappa$.
	
	Next, we consider the tails of Eq.~\eqref{eq:B25}. For $X \to \pm\infty$,
	$\Upsilon(X) = \left(\sqrt{1+\kappa^2} \pm \kappa\right)|X| + \order{1}$. We
	substitute into Eq.~\eqref{eq:B25} and use
	$\left(\sqrt{1+\kappa^2}+\kappa\right)\left(\sqrt{1+\kappa^2}-\kappa\right) = 1$ to
	obtain
	\begin{equation}
		\mathbb{E}[\lambda(\omega)] \sim q_{\pm}(\kappa)\,|\omega - \omega_C|^{-1/2},
		\label{eq:B27}
	\end{equation}
	where for each branch, $\scale_\pm(\freq)$ is defined as
	\begin{equation}
		q_{\pm}(\kappa) = \frac{\left(\sqrt{1+\kappa^2} \mp \kappa\right)^{1/2}}
		{(1+\kappa^2)^{1/2}},
	\end{equation}
	for $1 - \eta \ll \pm(\omega - \omega_C) \ll 1$. The factor $1-\eta$ cancels in
	Eq.~\eqref{eq:B27}, and the amplitudes coincide with those of Eq.~\eqref{eq:crit-asymp-scale}.
	Equation~\eqref{eq:B27} is therefore the overlap region in which the inner solution
	\eqref{eq:B25} matches the outer solution \eqref{eq:crit-asymp}
	\cite[Sec.~7.2]{bender1999advanced}.
	
	In summary, $\mathbb{E}[\lambda(\omega)]$ displays complete similarity in the
	distinguished limit \cite[Ch.~5, Sec.~5.1]{barenblatt1996scaling}. The product
	$(1-\eta)^{1/2}\,\mathbb{E}[\lambda(\omega)]$ depends on $\omega$ and $\eta$ only through
	$X$, and its behavior as $X \to \pm\infty$ is a pure power law. Consequently, for
	$\eta < 1$ the law $|\omega - \omega_C|^{-1/2}$ is an intermediate asymptotic, valid for
	$1 - \eta \ll |\omega - \omega_C| \ll 1$. For $|\omega - \omega_C| = \order{1-\eta}$,
	$\mathbb{E}[\lambda(\omega)]$ is bounded and given by Eq.~\eqref{eq:B25}.
	
	The exponent of the local behavior of NAPS at $\omega_C$ is thus not a continuous function of
	$\eta$. It is zero for every fixed $\eta < 1$ and $-1/2$ at $\eta \to 1^-$, where the
	inner region shrinks to the point $\omega_C$.
	
	The above derivation assumes a fixed $\kappa > 0$, so that the term of $\Lambda_1(\omega;\abxi)$ for $\sigma = +1$ is regular. For $\kappa = 0$, both terms in Eq.~\eqref{eq:A10} coincide and are singular, so the right-hand side of Eq.~\eqref{eq:B25} must be doubled. With
	$\Upsilon(X) = \sqrt{1+X^2} + 1$, the overlap region then gives $2\omega^{-1/2}$. This is
	consistent with the leading term of Eq.~\eqref{eq:kappa0_overlap}, since the distinguished scale
	$1 - \eta^2$ of Appendix~\ref{app:rho-0-close-omega-0} is asymptotically $2(1-\eta)$.

	\subsection{Behavior at $\omega\to\infty$}\label{app:naps-large-omega}
	
	In this section, we prove Lemma~\ref{lemma:naps_infty}. That is, we prove that for sufficiently large $|\omega| \to \infty$, $\ee[\lambda(\omega)]$ decays with a dominant term of $O(\omega^{-2})$. To achieve the above, we prove that 
	\begin{equation}
		\Lambda_1(\omega;\abxi) = \frac{1}{\omega^2} + O(\omega^{-3}),
	\end{equation}
	for sufficiently large $\omega$. Without loss of generality, we assume $\omega>0$. For all $\omega>\freq$, the absolute value in $A(\omega;\freq,\s)$ evaluates smoothly to $|\s\omega+\freq|=\omega+\freq\s$. Therefore, the function $H(\omega;\abxi)$ is given by:
	\begin{align*}
		H(\omega;\abxi) &= (\freq^2-1)\omega^2-4\s\freq\omega \\ 
		&\quad+ 1-\freq^2 -4\C^2,
	\end{align*}
	for all $\omega>\freq$, $\freq>0$. We now analyze the radicand of $\mathcal{K}(\omega;\abxi)$, that is,
	\begin{align}
		P(\omega;\abxi) & \triangleq H^2(\omega;\abxi ) + 4A^2(\omega;\freq,\s)B^2(\omega;\freq,\s), \notag
		\intertext{for all $\omega\in\rr$. Using algebraic manipulations, we can write $P(\omega;\abxi)$ as}
		&= [(\freq^2 + 1)\omega^2 +\freq^2+1-4\C^2]^{2} \notag \\
		&\quad+\C^2 A^2(\omega;\freq,\s),\notag
		\intertext{for all $\omega>\freq, $ $\freq>0$, $\proximity\in(0,1)$, and $\s\in\sdom$. We can now expand the resolvent in powers of $\omega$, to obtain}
		&= (\freq^2 + 1)^2 \omega^4 \notag\\ 
		&\quad + \left[ 2(\freq^2 + 1)^2 - 8\C^2(\freq^2 - 1) \right] \omega^2 \notag\\ 
		&\quad + 32 \C^2 \sigma \freq \omega + O(1),\notag
		\intertext{for all $\omega>\freq$, where $O(1)$ is a constant term that does not depend on $\omega$. Factoring out the term $(\freq^2+1)^2\omega^4$ and taking the square root via the binomial expansion $(1+u)^{1/2}=1+u/2+O(u^2)$ gives}
		\sqrt{P(\omega;\abxi)} &=(\freq^2 + 1)\omega^2 + \freq^2 + 1 \notag \\ 
		&\quad - \frac{4 \C^2(\freq^2 - 1)}{\freq^2 + 1}\notag 
		\\ 
		&\quad + \frac{16 \C^2 \sigma \freq}{(\freq^2 + 1)\omega} + O(\omega^{-2}), \label{eq:sqrt-p-large-om}
		\intertext{for sufficiently large $\omega>0$. We substitute Eq.~\eqref{eq:sqrt-p-large-om} into the definition of $\mathcal{K}(\omega;\abxi)$ yields}
		\mathcal{K}(\omega;\abxi) &= 2\omega^2  \Bigg[ 1 + \frac{2\sigma\freq}{\omega} + \frac{\freq^2}{\omega^2} \notag
		\\ 
		&\quad + \frac{4 \C^2}{(\freq^2 + 1)\omega^2} + O(\omega^{-3}) \Bigg]. \label{eq:k-large-om}
		\intertext{Applying the binomial expansion again to get the square root of $\mathcal{K}(\omega;\abxi)$, as expressed in Eq.~\eqref{eq:k-large-om}, yields}
		\sqrt{\mathcal{K}(\omega;\abxi)}  &= \sqrt{2}\omega \Bigg[ 1 + \frac{\s\freq}{\omega}  + \frac{2\C^2}{(\freq^2+1)\omega^2} + O(\omega^2)
		\Bigg],\label{eq:sqrt-k-large-om}
		\intertext{for all sufficiently large $\omega>0$. The final steps of this proof include dividing Eq.~\eqref{eq:sqrt-k-large-om} by $\sqrt{2}A(\omega;\abxi)$, for large $\omega>0$, and then substituting the result to Eq.~\eqref{eq:lambda-k}. The above steps result to }
		\Lambda_{1}(\omega;\abxi)  &= \frac{1 + \freq^2}{2 C^2(\proximity, \freq)} \left[ \frac{2 \C^2}{(\freq^2 + 1)\omega^2} + O(\omega^{-3}) \right] \notag
		\\ &= \frac{1}{\omega^2} + O(\omega^{-3}), \notag
	\end{align}
	for sufficiently large $\omega>0$. Since the steps do not assume a particular sign of $\s$, the result translates to the behavior of $\ee[\lambda(\omega)]$, at large $\omega$. As discussed previously, the same steps hold for $\omega<0$, due to the symmetry of $\ee[\lambda(\omega)]$ to the transformation $\omega\to-\omega$.

		\subsection{Simplification for $\freq=0$}\label{app:rho-0}
		
		When $\freq=0$, the network loses its rotational structure. In this appendix, we derive an exact equation for $\Lambda_1(\omega;\abxi)$, $\omega\in\rr$, and thus $\ee[\lambda(\omega)]$. For $\freq=0$,
		\begin{align}
			A(\omega;\freq,\s) = |\omega| \text{, }
			B(\omega;\freq,\s) = - 1 \text{, and }
			\C = \proximity/2. \label{eq:abc-rho0}
		\end{align}
		Substituting Eq.~\eqref{eq:abc-rho0} in the definition of $H(\omega;\abxi)$, for all $\omega\in\rr$, yields
		\begin{align}
			H(\omega;\abxi)=1-\omega^2-\proximity^2,
		\end{align}
		for all $\omega \in \rr$. 
		
		\subsubsection{Behavior close to $\omega\to0$}\label{app:rho-0-close-omega-0}
		
		Substituting into Eq.~\eqref{eq:lambda1-closed} gives
		\begin{align}
			\Lambda_1(\omega;\abxi)\bigg|_{\freq=0} &=\frac{\sqrt{2}}{\proximity^2|\omega|}  \Bigg[ \sqrt{(1-\omega^2-\proximity^2)^2 + 4\omega^2 } \notag \\
			&\qquad - (1-\omega^2-\proximity^2)
			\Bigg]^{1/2} - \frac{2}{\proximity^2}, \label{eq:lambda-rho-0}
			\intertext{for all $\omega \in\rr \setminus \{0\}$. As discussed in Appendix.~\ref{app:notation}, $\ee[\lambda(\omega)]$ is an even function. In this particular case, where $\freq=0$, in Eq.~\eqref{eq:lambda-rho-0}, $\sigma$ is not in the equation. Therefore, we assume $\omega>0$, without loss of generality. We can rationalize the numerator to get}
			&=\frac{2\sqrt{2}}{\proximity^2} \Bigg[ \sqrt{(1-\omega^2-\proximity^2)^2 + 4\omega^2 } \notag \\
			&\qquad + (1-\omega^2-\proximity^2)
			\Bigg]^{-1/2} - \frac{2}{\proximity^2}. \label{eq:rational-num-rho-0}
			\intertext{At $\omega\to0$ Eq.~\eqref{eq:rational-num-rho-0} behaves as }
			\lim_{\omega\to0}\Lambda_1(\omega;\abxi)\bigg|_{\freq=0}&=\frac{2}{\proximity^2\sqrt{1-\proximity^2}} - \frac{2}{\proximity^2}.\label{eq:rational-num-rho-0-lim}
		\end{align}
		Notice that Eq.~\eqref{eq:rational-num-rho-0-lim} matches the same as the generic formula for the limit as $\omega\to-\s\freq$~(see Eq.~\eqref{eq:lim-lambda-neg-sig-rho}), if we substitute $\freq=0$.
		To analyze the non-uniformity as $\proximity \to 1^-$ and $\omega \to 0$, we apply the method of matched asymptotics~\cite[see][Sec.~7.2]{bender1999advanced}, and also see Appendix~\ref{app:singularity-asympt-subcrit}. Define the small perturbation parameter: 
		\begin{equation}
			\delta = 1 - \proximity^2 \to 0^+ \quad (\text{so } \proximity^2 = 1 - \delta).
		\end{equation}
		When $\omega \to 0$ for a fixed $0 < \proximity < 1$, the regular expansion gives
		$$\Lambda_1(\omega;\abxi)\bigg|_{\freq=0} \sim \frac{2}{\proximity^2} \left( \frac{1}{\sqrt{\delta}} - 1 \right) - \frac{1}{\delta^{5/2}} \omega^2 + O(\omega^4),$$ 
		a result which follows from applying the binomial expansion twice on Eq.~\eqref{eq:rational-num-rho-0}. As $\delta \to 0^+$, the coefficients diverge ($\delta^{-1/2}$, $\delta^{-5/2}$), which indicates a singular boundary layer near $\omega \to 0$. The radicand inside can be written as 
		\begin{equation*}
			R(\omega; \delta) = \delta^2 + 4\omega^2 - 2\delta \omega^2 + \omega^4
		\end{equation*}
		Applying the method of dominant balance as $\delta \to 0^+$ and $\omega \to 0$: The leading balance between the terms is $\delta^2 \sim 4\omega^2$, giving the distinguished scale: 
		$$\omega = \delta \, X = (1 - \proximity^2) X$$ 
		where $X$ is an $O(1)$ term. Substituting $\omega = \delta X$ and $\proximity^2 = 1 - \delta$, $R(\omega)$, applying the binomial expansion once again, yields to the intermediate overlap of
		\begin{equation}
			\Lambda_1(\omega;\abxi) \sim \frac{2}{\sqrt{\omega}}-2, \label{eq:kappa0_overlap}
		\end{equation}
		for all $1-\proximity^2 \ll \omega \ll 1$, that is, when $\omega$ is much greater than $1-\proximity^2$, and $\omega\ll1$. At the same time, when $\omega \lesssim 1 - \proximity^2$, the region is too close to the smooth peak, where, as we discussed above, the peak of $\Lambda_1(\omega;\abxi)$ is flat~(see Eq.~\eqref{eq:rational-num-rho-0} and Eq.~\eqref{eq:rational-num-rho-0-lim}). Far from the peak, as we discuss here, $\Lambda_1(\omega;\abxi)$ shows a tail of order $\omega^{-2}$, that is, the same as the behavior for arbitrary $\freq>0$~(see Appendix~\ref{app:naps-large-omega}).

		\subsubsection{Behavior close to $\omega\to\infty$}
		
		The behavior of $\Lambda_1(\omega;\abxi)$ at large $\omega$ follows from the application the binomial theorem twice. Starting from Eq.~\eqref{eq:lambda-rho-0},
		\begin{align*}
			&=\frac{\sqrt{2}}{\proximity^2(1+\freq^2)\omega} \notag \\ 
			&\quad \times \Bigg[ \omega^2 \sqrt{ 1 + \frac{4-2(1-\proximity^2)}{\omega^2}+ \frac{(1-\proximity^2)^2}{\omega^4} } \notag \\
			&\qquad +\omega^2 \left(1 + \frac{\proximity^2-1}{\omega^2} \right)
			\Bigg]^{1/2} - \frac{2}{\proximity^2(1+\freq^2)},
			\intertext{for all $\omega>0$. Substituting for $\freq=0$ and applying the generalized binomial theorem yields}
			&=\frac{\sqrt{2}}{\proximity^2}\Bigg[ 1 + \frac{1}{2}\left( \frac{4-2(1-\proximity^2)}{\omega^2}+ \frac{(1-\proximity^2)^2}{\omega^4}\right) \notag 
			\\
			&\qquad -\frac{1}{8} \left( \frac{4-2(1-\proximity^2)}{\omega^2}+ \frac{(1-\proximity^2)^2}{\omega^4}\right)^{2} \notag \\
			&\qquad+ O(\omega^{-6})  + \left(1 + \frac{\proximity^2-1}{\omega^2} \right)
			\Bigg]^{1/2} - \frac{2}{\proximity^2},
			\intertext{for all $\omega>0$. Factoring out 2 from the outer radicand and applying the generalized binomial theorem once more}
			&=\frac{2}{\proximity^2} \Bigg[ 1 + \frac{\proximity^2}{2\omega^2} - \frac{4\proximity^2 + \proximity^4}{8\omega^4} + O(\omega^{-6}) \Bigg]  - \frac{2}{\proximity^2} \notag \\
			&=\frac{1}{\omega^2} - \frac{4 + \proximity^2}{4\omega^4} + O(\omega^{-6}),
		\end{align*}
		for all sufficiently large $|\omega|\gg1$. As discussed in the first paragraph of Appendix~\ref{app:rho-0-close-omega-0}, all above arguments hold for $\omega<0$.
		
		\section{Derivation of second moment $\mathbb{E}[\lambda^2(\omega)]$}\label{app:ED}
		
		\subsection{Derivation}
		
		In this appendix, we derive a closed-form expression for the second moment of the random SDM eigenvalues $\lambda(\omega)$, in the large-$N$ limit, as in Appendix~\ref{app:naps}. To achieve that, we work with the Eq.~\eqref{eq:Lambda-def}, and apply the Plancherel-Parseval identity, as in Eq.~\eqref{eq:Lambda-fourier-integ}. Specifically, we work with the integral
		\begin{align}
			\Lambda_2(\omega;\abxi) &= \frac{(1+\freq^2)^2}{2\pi} \int_{-2}^{2}\phi(u) \psi^2(u;\abxi)\, \dd u, \label{eq:integ-secon-mom}
			\intertext{for all $\omega \in \rr$, $\freq>0$, $\proximity\in[0,1)$, $\s\in\sdom$. Applying the Fourier transform on both spectral functions in the integral yields}
			&= \frac{(1+\freq^2)^2}{2\pi} \int_{-\infty}^{\infty} \mathcal{F}\phi(x) \mathcal{F}[\psi^2](x;\abxi)\, \dd x, \label{eq:lambda2-ft-def}
		\end{align}
		for all $\omega\in\rr$. To evaluate the Fourier transform of $\psi^2$, we invoke the convolution theorem~\cite[see][Sec.~4.4, Eq.~(4.56)]{Oppenheim1996}, which allows us to use $\mathcal{F}[\psi^2]=\mathcal{F}\psi \star \mathcal{F}\psi,$ where $\star$ denotes the convolution of two complex valued functions. The two-fold convolution of $\psi$, $\mathcal{F}[\psi^2]=(\mathcal{F} \psi)^{(2\star)}$ is equal to 
		\begin{align}
			&(\mathcal{F}\psi)^{(2\star)}(x; \abxi)\notag \\
			&\quad = \left( \frac{\pi |x|}{A^2(\omega; \freq, \s) \C ^2} + \frac{\pi}{2A^3(\omega; \freq, \s)} \right)
			e^{\mathcal{E}(x; \omega, \abxi)}, \label{eq:two-fold-h}
			\intertext{for all $x\in\rr$, $\freq >0$, $\proximity\in[0,1)$, $\s\in\sdom$, $\omega\in \rr \setminus\{-\s\freq\}$, where $\mathcal{E}$ is defined as}
			&\mathcal{E}(x; \omega, \abxi)=\frac{2\pi}{\C}\left[-A(\omega;\freq,\s) |x| +\iu B(\omega;\freq,\s) x \right]. \label{eq:chi}
		\end{align}
		From Eq.~\eqref{eq:two-fold-h}, we observe the following. First, the frequency $-\s\freq$ is also excluded, as in Eq.~\eqref{eq:lambda1-closed}. Second, the Fourier transform of $\psi$ is a part of the equation. This property foreshadows that the equation of $\Lambda_2(\omega;\abxi)$ will involve in some way $\Lambda_1(\omega;\abxi)$.
		
		Substituting Eq.~\eqref{eq:two-fold-h} into Eq.~\eqref{eq:lambda2-ft-def}, and describing the second term in terms of $\Lambda_1(\omega;\abxi)$ yields
		\begin{align}
			\Lambda_2(\omega;\abxi) &= \frac{(1+\freq^2)^2}{2\pi}  \frac{\pi}{A^2(\omega;\freq,\s)}  \notag \\
			&\quad \times \int_{-\infty}^{\infty}\sgn(x)e^{\mathcal{E}(x;\omega,\abxi)} J_1(4\pi x) \, \dd x \notag \\
			&\quad+\frac{1+\freq^2}{2A^2(\omega;\freq,\s)} \Lambda_1(\omega;\abxi), \label{eq:lambda2-integ-lambda1}
			\intertext{for all $\omega \in \rr \setminus \{-\s\freq\}$, where for all $x \in \rr$, $\mathcal{E}(x;\omega;\abxi)$ is as defined in Eq.~\eqref{eq:chi}, and $J_1$ is the Bessel function of the first kind and of order 1, as discussed in Eq.~\eqref{eq:fourier-semicirc}. Using~\cite[see][Sec.~9.11, Tab.~9.2]{Poularikas2000}, we derive the closed form formula of Eq.~\eqref{eq:lambda2-integ-lambda1}. Specifically,}
			&= \frac{2 \pi}{\proximity^2 A^2(\omega; \freq,\s)} \sum_{s \in\sdom} G(\omega;\abxi,s)
			\notag \\ 
			&\quad + \frac{1+\freq^2}{2A^2(\omega;\freq,\s)}\Lambda_1(\omega;\abxi), \label{eq:l2-closed}
		\end{align}
		for all $\omega \in \rr \setminus \{-\s\freq\}$, where
		\begin{align}
			G(\omega;\abxi,s) &= \frac{\sqrt{(4\pi)^2+F^2(\omega;\abxi,s)} -F(\omega;\abxi,s)}{4\pi\sqrt{(4\pi)^2+F^2(\omega;\abxi,s)}} 
			\intertext{and }
			F(\omega;\abxi,s)&=\frac{2\pi }{\C} [A(\omega;\freq,\s) + \iu s B(\omega;\freq,\s)], \label{eq:F}
			\intertext{for all $\omega \in \rr \setminus \{-\s\freq\}$. Notice that $F$ relates to $\mathcal{E}$ via the equation}
			F(\omega;\abxi,s)&=-\mathcal{E}(-s;\omega;\abxi), \notag
		\end{align}
		for all $\omega \in \rr \setminus \{-\s\freq\}$, where $s \in \sdom$ is a sign variable, which $G(\omega;\abxi, s)$ in Eq.~\eqref{eq:l2-closed} is summed over.
		
		\subsection{Connection to normalized effective dimensionality}
		
		In the large-$N$ limit, normalized effective dimensionality is defined as
		\begin{align}
			\lim_{N\to\infty} D_{N}(\omega) &= \frac{\ee[\lambda(\omega)]^2}{\ee[\lambda^2(\omega)]} \notag
			\\
			&= \frac{\left( \sum_{\s \in\sdom } \Lambda_1(\omega;\abxi)\right)^2}{ \sum_{\s \in\sdom } \Lambda_{2}(\omega;\abxi)},
			\label{eq:ed-lambda}
		\end{align}
		for all $\omega\in\rr \setminus \{-\freq, \freq\}$.
		
		\subsection{Behavior close to $\omega_C$}
		\label{app:C3}
		
		In this appendix, we prove Lemma~\ref{lemma:dim-collapse}. Since both moments are even
		functions of $\omega$, so is $\lim_{N\to\infty} D_N(\omega)$, and we assume
		$\omega > 0$ without loss of generality. 
		
		We start from the closed form expression for the second moment per branch, as given by Eq.~\eqref{eq:l2-closed}, and apply the following steps. First, we expand the function $F$ at $\eta \to 1^-$ and express $\omega$ as $\omega=\freq+\varepsilon$. Then, $\C=(1+\freq^2)/2$, $A(\omega;\freq,\s)=|\varepsilon|$, and $B(\omega;\freq,\s)=-(1+\freq^2)-\freq\varepsilon$. Hence,
		\begin{align}
			F(\freq+\varepsilon;\abxi,s) =- 4\pi \iu s + \frac{4\pi}{1+\freq^2}(|\varepsilon| - \iu s \freq \varepsilon). \label{eq:F-eps}
		\end{align}
		Then, using Eq.~\eqref{eq:F-eps}, the radicand of the function $G$ can be expressed as 
		\begin{align}
			(4\pi)^2 + F^2(\freq+\varepsilon;\abxi,s) = -\frac{32\pi^2}{1+\freq^2}(|\varepsilon| - \iu s \freq \varepsilon) + \order{\varepsilon^2}, \label{eq:G-radicand-asymp}
		\end{align}
		for all $\varepsilon \in \rr$, $\freq >0$, and $s \in \sdom$. Equation~\eqref{eq:G-radicand-asymp} allows for the sum of $G$ over $s\in\sdom$ to be expressed as
		\begin{align}
			&\sum_{s\in\sdom} G(\freq+\varepsilon;\abxi,s) 
			\notag \\ 
			&\quad= -\frac{\left(\sqrt{1+\freq^2} + \freq\, \sgn(\varepsilon) \right)^{1/2}}{4\pi |\varepsilon|^{1/2}} + \order{1},
		\end{align}
		for all $\varepsilon \in \rr$ and $\freq >0$. From Lemma~\ref{lemma:naps_singular}, at $\proximity\to 1^{-}$, $\Lambda_{1}(\freq+\varepsilon; \abxi) \sim 2\scale_{\pm}(\freq) |\varepsilon|^{-1/2}$. Substituting into Eq.~\eqref{eq:l2-closed} yields
		\begin{align}
			\Lambda_{2}(\freq+\varepsilon;\abxi) &\sim |\varepsilon|^{-5/2} \Big[ \sqrt{1+\freq^2}\left( \sqrt{1+\freq^2} - \freq\, \sgn(\varepsilon)\right)^{1/2} \notag 
			\\ &\quad- \frac{1}{2}\left( \sqrt{1+\freq^2} - \freq\, \sgn(\varepsilon)\right)^{1/2}\Big] \notag
			\\
			&=\frac{1}{2}\left( \sqrt{1+\freq^2} - \freq \, \sgn(\varepsilon)\right)^{3/2} |\varepsilon|^{-5/2}, \label{eq:L2-asympt}
		\end{align}
		for all $\varepsilon \in \rr$ sufficiently close to 0, $\freq>0$, and $\eta \to 1^-$. Substituting Eq.~\eqref{eq:L2-asympt} to Eq.~\eqref{eq:ed-lambda} yields
		\begin{align}
			\lim_{N\to \infty} D_N(\omega) &= \frac{\left( \sum_{\s \in\sdom } \Lambda_1(\omega;\abxi)\right)^2}{ \sum_{\s \in\sdom } \Lambda_{2}(\omega;\abxi)} \notag \\
			&\sim \frac{4\left( \sqrt{1+\freq^2} \pm \freq\right)^{1/2}}{1+\freq^2}|\omega-\omega_C|^{3/2},
			\intertext{or, more compactly,}
			\lim_{N\to \infty} D_N(\omega)&\sim d_\pm(\freq)|\omega-\omega_C|^{3/2},
		\end{align}
		as $\omega \to \omega_C$, where 
		\begin{equation}
			d_\pm(\freq) \triangleq  \frac{4\left( \sqrt{1+\freq^2} \pm \freq\right)^{1/2}}{1+\freq^2}. \label{eq:d-pm}
		\end{equation}
		In Eq.~\eqref{eq:d-pm}, the positive sign corresponds to $|\omega|>|\omega_C|$, and the negative sign to $|\omega|<|\omega_C|$.
		
		\section{Derivation of first eigenvalues moment in the time domain $\mathbb{E}[\lambda(\tau)]$}
		\label{app:first-moment-eigenvalues-time}
		
		\subsection{Subcritical regime: $\proximity<1$}\label{app:naac-subcritical}
		
		\subsubsection{Exact formula}
		In this appendix, we derive the closed formula of NAAC in the large-$N$ limit. That is,
		\begin{equation}
			\lim_{N\to\infty}\Tr_N[\mathbf{C}(\tau)] = \frac{1}{2\pi}\int_{-\infty}^{\infty}\mathbb{E}[\lambda(\omega)] e^{\iu\omega \tau} \,  \dd\omega,
			\label{eq:naac-naps-relation}
		\end{equation}
		for all $\tau\in\mathbb{R}$, which we will abbreviate as $\mathbb{E}[\lambda(\tau)]$. This is the \emph{inverse Fourier transform} over the cyclical frequency $\omega\in\mathbb{R}$. To calculate this integral, we start from taking the inverse Fourier transform of Eq.~\eqref{eq:h-integral}. That is,
		\begin{align}
			\mathbb{E}[\lambda(\tau)] &= \frac{1}{4\pi} \sum_{\sigma\in\{-1,1 \}} \int_{-\infty}^{\infty} e^{\iu\omega\tau}  \notag \\
			&\quad\times  \int_{-2}^{2}\frac{\sqrt{4 - u^2}}{(1-\proximity u/2)^2 + (\omega + \sigma \proximity\freq u/2)^2} \, \dd u \, \dd\omega, \notag
			\intertext{for all $\tau\in\mathbb{R}$. We can swap the order of the integrals to obtain}
			&= \frac{1}{4\pi} \sum_{\sigma\in\{-1,1 \}} \int_{-\infty}^{\infty} \sqrt{4 - u^2} \notag \\
			&\quad\times  \int_{-2}^{2} \frac{e^{\iu\omega\tau}}{(1-\proximity u/2)^2 + (\omega + \sigma \proximity\freq u/2)^2} \, \dd \omega \, \dd u
			\intertext{The inner integral is the inverse Fourier transform (with the exponent of $\iu \omega$) of the {Lorentzian} $\psi$ function~(see Eq.~\eqref{eq:phi-psi}), this time with respect to $\omega$. Luckily, $\psi$ can be treated in a very similar way~\cite{campbell_fourier_1948}, resulting to }
			&= \frac{1}{4\pi} \sum_{\sigma\in\{-1,1 \}} \int_{-2}^{2} \sqrt{4-u^2} \notag \\
			&\quad  \times\frac{1}{2|1 - \proximity u/2|} e^{-|1-\proximity u/2||\tau|}e^{-\iu \sigma \proximity \freq u\tau/2} \dd u, \notag
			\intertext{for all $\tau\in\mathbb{R}$. Notice that $1-\proximity u/2 \geq0$, for all $\proximity\in[0,1)$, $u \in [-2,2]$. Also, we apply the substitution: $w=(u/2 - 1)/2$, such that $w\in[0,1]$. This way, the integral takes the following form}
			&= \frac{2e^{-(1+\proximity)|\tau|}}{\pi} \sum_{\s \in\sdom} {e^{\iu \s \proximity\freq \tau}} \notag 
			\\
			&\quad \times \int_{0}^{1} \frac{w^{1/2}(1-w)^{1/2}}{1+\proximity-2\proximity w} e^{2\proximity|\tau| w} e^{-2 \iu \s \proximity \freq \tau  w} \, \dd w,\label{eq:integ-phi}
			\intertext{for all $\tau\in\mathbb{R}$. The integral in Eq.~\eqref{eq:integ-phi} can be expressed in terms of the confluent hypergeometric function of two variables $\Phi_1$ \cite[see][Chap. IV]{Arshad2024-oa}. By following algebraic steps yields}
			&= \frac{ e^{-(1+\proximity)|\tau|}}{4(1+\proximity)} \sum_{\s \in\sdom} e^{\iu \s \proximity\freq \tau} \notag 
			\notag \\
			&\quad \times \Phi_1\left(\frac{3}{2},1,3;\frac{2\proximity}{1+\proximity}, 2\proximity|\tau|-2\iu \s\proximity\freq\tau \right), \label{eq:exact-phi1-app}
		\end{align}
		for all $\tau \in \rr$.

		\subsubsection{Asymptotic expansion}
		In addition, for sufficiently large $|\tau|$, the following asymptotic expansion holds, following from \cite[see][Ch.~5, Sec.~5.11]{Erdelyi1953-au}:
		\begin{align}
			&\ee[\lambda(\tau)]\bigg|_{\proximity<1} 
			= \frac{2 e^{-(1-\proximity)|\tau|}}{\sqrt{\pi}(1-\proximity)(1+\freq^2)^{3/4}(2\proximity|\tau|)^{3/2}}
			\notag \\ &\quad \cos\left(\freq\proximity |\tau|-\frac{3}{2}\arctan\freq \right) + O\left( \frac{e^{-(1-\proximity)|\tau|}}{|\tau|^{5/2}}\right)\label{eq:subcrit-asymp}
		\end{align}
		for all $\freq>0$, $\proximity \in (0,1)$.
		
		\subsection{Critical regime: $\proximity\to1^{-}$}\label{app:naac-critical}

		\subsubsection{Exact formula}
		
		Although it is possible to derive the simplification of $\ee[\lambda(\omega)]$ for $\proximity\to1^-$, starting from Eq.~\eqref{eq:exact-phi1-app}, via the relation between Humbert's confluent hypergeometric function of two variables $\Phi_1$ and Kummer's confluent hypergeometric function~\cite{buchholz_confluent_1969}, we present a simpler, equivalent way to get to the simplified formula at criticality.
		
		Specifically, we start from Eq.~\eqref{eq:integ-phi} and take the limit as $\proximity \to 1^-$. This makes the calculation of the integral equivalent to taking the Laplace transform of $w^{1/2}(1-w)^{-1/2}$ over the interval $[0,1]$~(see Lemma~\ref{lemm:lap-tr-sq} below). Then, the rest follows from algebraic manipulations.
		
		\begin{lemma}\label{lemm:lap-tr-sq}
			Consider the function $\chi(w)=w^{1/2}(1-w)^{-1/2}$, for all $w\in[0,1)$ and zero outside. The Laplace transform of $\chi$ is equal to
			\begin{equation}
				\mathcal{L} \chi (p) = \frac{\pi}{2} e^{-p/2} \left[ I_0\left(\frac{p}{2}\right) - I_1\left(\frac{p}{2}\right) \right],
			\end{equation}
			for all $p \in \mathbb{C}$, where $I_0, I_1$ are the modified Bessel functions of the first kind of order 0 and 1, respectively.
		\end{lemma}
		\begin{proof}
			We cite the following Laplace transform pair~\cite[see][Sec.~4.3, Eq.~(14)]{Erdelyi1954TablesVol1}. For a real function $f_0$ defined as
			\begin{align*}
				f_0(t) = \begin{cases} (2bt - t^2)^{\nu - 1/2}, & 0 < t < 2b \\ 0, & t > 2b \end{cases}, 
			\end{align*}
			where $\nu>-1/2$, the Laplace transform over $[0,\infty)$ is equal to 
			\begin{align*}
				g_0(p) = \pi^{1/2} \Gamma\left(\nu + \frac{1}{2}\right) \left(\frac{2b}{p}\right)^\nu e^{-bp} I_\nu(bp),
			\end{align*}
			for all $p \in \mathbb{C} \setminus \{ 0\}$. Setting \(b = \frac{1}{2}\) and \(\nu = 0\) yields
			\begin{align*}
				\mathcal{L}\left\{ \frac{1}{\sqrt{w(1-w)}} \right\}(p) = \pi e^{-p/2} I_0\left(\frac{p}{2}\right),
			\end{align*}
			for all $p \in \mathbb{C}$. Applying the operational multiplication property~\cite[see][Sec.~4.1, Eq.~(6)]{Erdelyi1954TablesVol1} yields
			\[ w^{1/2}(1-w)^{-1/2} = \frac{w}{\sqrt{w(1-w)}} = w f_0(w). \]
			Differentiating with respect to \(p\) and using the Bessel relation \(I_0'(z) = I_1(z)\) gives
			\begin{align*}
				\mathcal{L}\left\{ w^{1/2}(1-w)^{-1/2} \right\}(p) &= -\frac{d}{dp}\left[ \pi e^{-p/2} I_0\left(\frac{p}{2}\right) \right] \\ &= \frac{\pi}{2} e^{-p/2} \left[ I_0\left(\frac{p}{2}\right) - I_1\left(\frac{p}{2}\right) \right],
			\end{align*}
			for all $p \in \mathbb{C}.$
		\end{proof}
		
		To prove Theorem~\ref{thr:naac-crit}, we start from Eq.~\eqref{eq:integ-phi}, and take its limit for $\proximity\to1^{-}$. This yields to
		\begin{align*}
			\ee[\lambda(\tau)]\bigg|_{\proximity\to1^-} &= \frac{2e^{2|\tau|}}{\pi} \sum_{\s \in\sdom} e^{\iu \s \freq \tau} \notag 
			\\
			&\quad \times \int_{0}^{1} w^{1/2}(1-w)^{-1/2} e^{2|\tau| w} e^{-2 \iu \s \freq \tau  w} \, \dd w,
			\intertext{for all $\tau \in \rr$. Invoking Lemma~\ref{lemm:lap-tr-sq} shown above gives}
			&=\frac{e^{-|\tau|}}{2}\sum_{\s \in\sdom} \notag  
			\\ 
			&\quad \times \left[ I_0\left(|\tau|-\iu\s\freq\tau\right) + I_1\left(|\tau|-\iu\s\freq\tau\right) \right],
		\end{align*}
		for all $\tau \in \rr.$
		
		\subsubsection{Asymptotic expansion}\label{app:asympt-crit}
		
		To derive the leading-order asymptotic expansion of
		\begin{align}
			\ee[\lambda(\tau)]\bigg|_{\proximity\to1^-} \notag &=\frac{e^{-|\tau|}}{2}\sum_{\s \in\sdom}\notag  
			\\ 
			&\quad \times \left[ I_0\left(|\tau|-\iu\s\freq\tau\right) + I_1\left(|\tau|-\iu\s\freq\tau\right) \right],\label{eq:eq-ee-l-tau-eta-1}
		\end{align}
		for large $|\tau|\gg1$, we use the classical asymptotic expansion for the modified Bessel functions of the first kind $I_\nu(z)$ as $|z| \to \infty$. For fixed order $\nu$ and $|z| \to \infty$ in the principal sector $|\arg z| < {\pi}/{2}$, the standard asymptotic expansion is~\cite[see][Sec.~7.13.1, Eq.~(5)]{Erdelyi1953Higher2}
		\begin{align*}
			I_\nu(z) &\sim \frac{e^z}{\sqrt{2\pi z}} \left( 1 - \frac{4\nu^2 - 1}{8z} + O\left(\frac{1}{z^2}\right) \right),
		\end{align*}
		for all $z\in\mathbb{C}$, such that $|\arg z|<{\pi }/{2}$. The orders we are interested in are $\nu=0,1$. Summing for both functions gives
		\begin{align}
			&I_0(z) + I_1(z) \sim \frac{2 e^z}{\sqrt{2\pi z}} \left( 1 - \frac{1}{8z} + O\left(\frac{1}{z^2}\right) \right), \label{eq:asympti0-i1} 
		\end{align}
		for all $z\in\mathbb{C}$, such that $|\arg z|<{\pi }/{2}$. The argument $z$ in Eq.~\eqref{eq:asympti0-i1} is substituted by $|\tau|-\iu \s\freq\tau$~(see Eq.~\eqref{eq:eq-ee-l-tau-eta-1}), for sufficiently large $\tau$. Let $z_\sigma = |\tau|(1 - \iu \sigma \freq \operatorname{sgn}(\tau))$. Since $\freq > 0$, we have $|\arg z_\sigma| = \arctan(\freq) < {\pi}/{2}$, placing $z_\sigma$ strictly inside the principal half-plane. Notice that $e^{z_\sigma} = e^{|\tau|} e^{-\iu \sigma \freq \tau}$. When multiplied by the prefactor ${e^{-|\tau|}}/{2}$, the real exponential factor cancels:
		$$ \frac{e^{-|\tau|}}{2} \left[ I_0(z_\sigma) + I_1(z_\sigma) \right] \sim \frac{e^{-\iu \sigma \freq \tau}}{\sqrt{2\pi z_\sigma}} \left( 1 - \frac{1}{8 z_\sigma} \right). $$
		In polar form, $w_\sigma = 1 - \iu \sigma \freq \operatorname{sgn}(\tau) = \sqrt{1+\freq^2} \, e^{-\iu \sigma \operatorname{sgn}(\tau) \arctan(\freq)}$. Taking the square root yields
		$$ \sqrt{z_\sigma} = \sqrt{|\tau|} (1+\freq^2)^{1/4} e^{-\iu \frac{1}{2} \sigma \operatorname{sgn}(\tau) \arctan(\freq)}, $$
		for all sufficiently large $|\tau|$. Combining with the numerator $e^{-\iu \sigma \freq \tau}$ and summing over the sign $\sigma$ results to
		\begin{align}
			&\sum_{\sigma \in \{-1,1\}} \frac{e^{-\iu \sigma \freq \tau}}{\sqrt{z_\sigma}}= \frac{1}{\sqrt{|\tau|} (1+\freq^2)^{1/4}} \\ 
			&\quad\times \sum_{\sigma \in \{-1,1\}} \exp\left( -\iu \sigma \left( \freq \tau - \frac{1}{2} {\sgn}(\tau) \arctan(\freq) \right) \right), \label{eq:comp-exp-i}
		\end{align}
		for all sufficiently large $|\tau|$. Since $\cos(-x) = \cos(x)$, the sign function $\operatorname{sgn}(\tau)$ simplifies, leaving purely even dependence on $|\tau|$. In addition, Euler's formula can be used to turn the complex exponential term in Eq.~\eqref{eq:comp-exp-i} in terms of the cosine function of real argument, 
		\begin{align}
			\sum_{\sigma \in \{-1,1\}} \frac{e^{-\iu \sigma \freq \tau}}{\sqrt{z_\sigma}} = \frac{2 \cos\left( \freq |\tau| - 0.5 \arctan \freq \right)}{\sqrt{|\tau|} (1+\freq^2)^{1/4}}. \label{eq:pre-final-i-eta-1-asymp}
		\end{align}
		Finally, substituting Eq.~\eqref{eq:pre-final-i-eta-1-asymp} back in Eq.~\eqref{eq:eq-ee-l-tau-eta-1}, yields
		\begin{align}
			\ee[\lambda(\tau)]\bigg|_{\proximity\to1^{-}} &= \sqrt{\frac{2}{\pi |\tau|}} \Bigg[ \frac{\cos\left( \freq |\tau| - \frac{1}{2} \arctan \freq \right)}{(1+\freq^2)^{1/4}} \notag \\ 
			&\quad - \frac{\cos\left( \freq |\tau| - \frac{3}{2} \arctan \freq \right)}{8 |\tau| (1+\freq^2)^{3/4}} \Bigg] \notag \\
			&\quad+ O\left( |\tau|^{-5/2} \right), 
			\label{eq:D12}
		\end{align}
		for all sufficiently large $|\tau|\gg1$.
		
		\subsubsection{Fourier duality at the edge of instability}\label{app:D2c}
		
		By Eq.~\eqref{eq:naac-naps-relation}, NAPS is the Fourier transform of NAAC,
		\begin{equation}
			\ee[\lambda(\omega)]
			= \int_{-\infty}^{\infty} \lim_{N\to\infty}\mathrm{Tr}_N[\mathbf{C}(\tau)]\,
			e^{-\mathrm{i}\omega\tau}\,d\tau .
		\end{equation}
		We show that the singular part of NAPS near $\omega_C$ comes only from the
		leading term of Eq.~\eqref{eq:C-critical-decay}.
		Fix $T>0$ and split the integral into the parts $|\tau|\le T$ and $|\tau|>T$.
		NAAC is bounded on $|\tau|\le T$, so the first part is a continuous function of
		$\omega$. On $|\tau|>T$, we write NAAC as the leading term of
		Eq.~\eqref{eq:C-critical-decay} plus a remainder. By Eq.~\eqref{eq:D12}, the
		remainder is $O(|\tau|^{-3/2})$. It is therefore absolutely integrable and also
		gives a continuous function of $\omega$. Both contributions are bounded near
		$\omega_C$, so only the leading term can produce a singularity.
		
		Since NAAC is even in $\tau$, the contribution of the leading term is
		\begin{align}
			&\leading = 2\sqrt{\frac{2}{\pi}}\,\left(1+\freq^2\right)^{-1/4}
			\notag \\
			&\quad \times\int_T^{\infty}\tau^{-1/2}
			\cos\!\left(\freq\tau-\frac{1}{2}\arctan\freq\right)\cos(\omega\tau)\,d\tau,
		\end{align}
		for all $\omega\in\rr$. The product-to-sum identity splits the integrand into a term with frequency
		$\freq+\omega$ and a term with frequency $\freq-\omega$. Near $\omega=\freq$,
		the first term stays bounded. The singular part of $\ee[\lambda(\omega)]$ near
		$\omega_C=\freq$ is therefore
		\begin{align}
			&\leading = \sqrt{\frac{2}{\pi}}\,\left(1+\freq^2\right)^{-1/4}\notag\\
			&\quad\times\int_T^{\infty}\tau^{-1/2}
			\cos\!\left((\freq-\omega)\tau-\tfrac{1}{2}\arctan\freq\right)d\tau .
		\end{align}
		For $a\neq 0$ and real $c$,
		\begin{equation}
			\int_0^{\infty} t^{-1/2}\cos(at-c)\,dt
			= \sqrt{\frac{\pi}{2|a|}}\,\bigl[\cos c+\mathrm{sgn}(a)\sin c\bigr].
		\end{equation}
		The integral over $[0,T]$ is bounded, so it does not change the singular part.
		We set $\omega=\freq+\varepsilon$, so that $a=\freq-\omega=-\varepsilon$.
		The factors $\sqrt{2/\pi}$ and $\sqrt{\pi/2}$ cancel, leading to
		\begin{align}
			&\ee[\lambda(\freq+\varepsilon)]
			\sim \left(1+\freq^2\right)^{-1/4}|\varepsilon|^{-1/2} \notag
			\\
			&\quad\times
			\left[\cos\!\left(\frac{1}{2}\arctan\freq\right)
			-\sgn(\varepsilon)\sin\!\left(\frac{1}{2}\arctan\freq\right)\right]
			, \label{eq:asymp-lambd-eps-fourier}
		\end{align}
		for sufficiently small $\varepsilon\in\rr$. The case $\varepsilon>0$, that is $|\omega|>|\omega_C|$, gives the minus sign inside the bracket. The case $\varepsilon<0$ gives the plus sign. The amplitude in Eq.~\eqref{eq:asymp-lambd-eps-fourier} equal to $\scale_{+}(\freq)$ for $|\omega|>|\omega_C|$ and $\scale_{-}(\freq)$ for $|\omega|<|\omega_C|$. The above holds because of the following algebraic identity
		\begin{equation}
			\left[\cos\!\left(\frac{1}{2}\arctan\freq\right)
			\mp\sin\!\left(\frac{1}{2}\arctan\freq\right)\right]^2
			= \frac{\sqrt{1+\freq^2}\mp\freq}{\sqrt{1+\freq^2}}, \label{eq:D18}
		\end{equation}
		for all $\freq\in\rr$. Substituting Eq.~\eqref{eq:D18} into Eq.~\eqref{eq:asymp-lambd-eps-fourier} yields Eq.~\eqref{eq:q-scale}. This agrees with Lemma~\ref{lemma:naps_singular}.

		Finally, $\ee[\lambda(\omega)]$ is even in $\omega$. Equation~\eqref{eq:asymp-lambd-eps-fourier} therefore holds at $\omega_C=-\freq$.

	\end{document}